\documentclass[twocolumn,amssymb, amsmath, aps, superscriptaddress, showpacs, footinbib, prb]{revtex4-1}

\usepackage{graphicx}
\usepackage{dcolumn}
\usepackage{bm}
\usepackage{placeins} 
\usepackage[per-mode=symbol]{siunitx}

\begin{document}


\title{ Signatures of quantum criticality in hole-doped and chemically pressurized EuFe$_2$As$_2$ single crystals }

\author{Jannis Maiwald}
\email{jmaiwal@gwdg.de}

\author{H.\,S. Jeevan}

\author{Philipp Gegenwart}
\affiliation{I. Physikalisches Institut, Georg-August-Universitaet Goettingen, D-37077
Goettingen, Germany}

\begin{abstract}
We study the effect of hole-doping and chemical pressure (isovalent doping) in single crystals of K$_x$Eu$_{1-x}$Fe$_2$As$_2$ and EuFe$_2$(As$_{1-y}$P$_y$)$_2$, respectively, by measurements of the thermopower, $S(T)$, and electrical resistivity, $\rho(T)$. The evolution of $S(T)$ upon doping indicates drastic changes of the electronic configuration at critical values $x_{\mathrm{cr}}=0.3$ and $y_{\mathrm{cr}}=0.21$, respectively, as the spin-density-wave transition is completely suppressed and superconductivity (SC) emerges. For the case of chemical pressure, the comparison with published ARPES measurements indicates a Lifshitz transition at $y_{cr}$. The temperature dependences $S(T)/T\propto \log T$ and $\Delta\rho\propto T$ observed in the normal state above the SC transition suggest quantum criticality in both systems.
\end{abstract}

\pacs{05.30.Rt, 74.70.Xa, 74.20.Mn, 74.25.fg}

\maketitle


\section{Introduction}

Unconventional superconductivity (SC) often occurs in systems with multiple competing phases and fluctuations.\cite{Norman11} In particular it has been demonstrated for heavy-fermion metals, that SC is maximal at the point, where long-range magnetic order disappears.\cite{Mathur98} Such a quantum critical point (QCP) leads to strong magnetic fluctuations in its vicinity in the phase diagram, which may act as glue for unconventional SC. It is therefore very important, to characterize quantum critical behavior in the normal state of unconventional superconductors.

In this paper, we focus on a possible QCP in the vicinity of the SC phase of the iron-pnictide EuFe$_2$As$_2$.\cite{Jeevan08b} This system belongs to the so-called ``122''-family of the recently discovered iron pnictides: $M$Fe$_2$As$_2$ ($M=$ divalent alkaline earth or rare earth metal: Ca, Sr, Ba and Eu). In order to induce SC in this compound, the spin-density-wave (SDW) type antiferromagnetic
(AF) ordering of FeAs layers with $T_{\mathrm{SDW}} = \SI{189}{\kelvin}$ needs to be suppressed. This can generally be achieved in two ways: application of external pressure or by chemical substitution of either K or Na to the Eu site (hole doping) or P to the As site (isovalent substitution).\cite{Terashima09b,Kurita11,Jeevan08b,Qi08, Ren09} The suppression of the AF order occurs continuously, thus the system is a candidate for a potential QCP. Due to the localized 4$f$-moments of the Eu atoms the system is the only ``122''-compound that features a second magnetic sublattice. The corresponding Eu$^{2+}$ moments order AF around $\SI{19}{\kelvin}$. However, the coupling between the two magnetic sublattices is weak.\cite{Jeevan08,Herrero09}

The 122-type iron pnictides feature a complex fermiology, consisting of three hole-like Fermi surfaces ($h$-FS) at the center of the first Brillouin zone [${\bf\Gamma}$-point; ${\bf k}=(0,0)$] and two electron-like Fermi surfaces ($e$-FS) at the corner [${\bf M}$-point; ${\bf k}=(\pi, \pi)$]. The two $e$-FSs and the two inner $h$-FSs are basically two-dimensional, while the outer $h$-FS has a more three-dimensional (3D) character due to a stronger dispersion along ${\bf k}_z$.\cite{Wen11, Johnston10, Paglione10} The SDW ordering takes place along the nesting vector ${\bf Q}_n$ in the ${\bf\Gamma}-{\bf M}$ direction. Upon hole doping, the $h$-FSs expand, while the $e$-FSs shrink along the $k_xk_y$-plane. This leads to a weakening of the nesting condition and hence to a suppression of the SDW ordering. On the other hand, the isovalent substitution of As with the smaller P leads to a substantial reduction of the unit-cell volume. Recent \emph{angle-resolved photoemission spectroscopy} (ARPES) measurements on EuFe$_2$(As$_{1-y}$P$_y$)$_2$ showed that the $e$-FSs stay almost unchanged, while the $h$-FSs become more 3D, because P substitution leads to a drastic reduction of the $c/a$ ratio,\cite{Jeevan10} enhancing the dispersion along $k_z$.\cite{Thirupathaiah11} The stronger warping of the $h$-FSs breaks the nesting condition and eventually leads to SC.\cite{Wu11}

In the case of iron-pnictides, a possible QCP is
hidden by SC and therefore indications of quantum critical behavior are
limited to temperatures above $T_\mathrm{c}\approx \SI{35}{\kelvin}$. Since the specific heat at this temperatures is highly dominated by the phonon contribution and (in our case) the
contribution of Eu$^{2+}$ moments, it is not possible to analyze non-Fermi liquid (NFL)
behavior in this property. Instead, we focus on the thermoelectric power,
S(T), which has recently been shown to be a very sensitive probe of quantum
criticality.\cite{Paul01} While thermopower divided by temperature, $S(T)/T$, is
constant for a Landau Fermi liquid, a logarithmic divergence, $S/T\propto \log
T$, is expected in the quantum critical regime for a two-dimensional SDW QCP.\cite{Paul01}
Note, that these temperature dependences are similar to those expected for
the specific heat coefficient $C/T$. Furthermore, we analyze the temperature
dependence of the electrical resistivity, for which the same model predicts
a linear temperature dependence,\cite{Kondo02} in sharp contrast to the $T^2$ behavior
characteristic for Landau Fermi liquids.

The paper is organized as follows: section \ref{methods} provides information on the
single crystal synthesis and characterization, while thermopower and
electrical resistivity results on K$_x$Eu$_{1-x}$Fe$_2$As$_2$ and EuFe$_2$(As$_{1-y}$P$_y$)$_2$ are
discussed in \ref{k-dope} and \ref{p-dope}, respectively. This is the first study on single
crystals of the former hole-doped system, while the phase diagram for
isovalent P-substitution has already been determined on single crystals
previously.\cite{Jeevan10, Zapf11}


\section{Methods} \label{methods}
A series of K$_x$Eu$_{1-x}$Fe$_2$As$_2$ single crystals was synthesized using the self-flux method, in which the crystals grow out of an FeAs flux. In a first step precursors of FeAs and KAs were synthesized using simple solid state reactions. Stoichiometric amounts of the respective starting elements (Fe \SI{99.998}{\percent}, As \SI{99.999}{\percent}, K \SI{99.95}{\percent}) were taken into an Al$_2$O$_3$ crucible, which was then sealed in a Ta crucible under argon atmosphere. The FeAs batch was heated to \SI{600}{\celsius} at a rate of \SI{100}{\celsius \per \hour} for \SI{10}{\hour} followed by a temperature of \SI{700}{\celsius} at a rate of \SI{50}{\celsius \per \hour} for \SI{20}{\hour} and was then quickly (\SI{250}{\celsius \per \hour}) cooled to room temperature. The KAs batch was slowly heated to \SI{700}{\celsius} at a rate of \SI{25}{\celsius \per \hour} and kept there for \SI{6}{\hour} after which it was also cooled to room temperature quickly.

Stoichiometric amounts (as for a 1:4:4 compound) of FeAs, KAs and Fe powder were thoroughly mixed with small pieces of Eu (\SI{99.99}{\percent}) and then sealed as described above. All steps had to be carried out under argon atmosphere to prevent the highly reactive materials KAs and Eu from reacting with the air. The single crystals of K$_x$Eu$_{1-x}$Fe$_2$As$_2$ were then grown by heating the batches by various temperature profiles. The most successful one regarding composition and size of the crystals was as follows: heating at a rate of \SI{50}{\celsius \per \hour} to \SI{1100}{\celsius}, staying a total of approximately \SI{4}{\hour} at this temperature, followed by the actual growth by decreasing the temperature at a rate of \SI{3}{\celsius \per \hour} down to \SI{900}{\celsius}. Afterwards the temperature was quickly reduced to room temperature. Using this method we were able to grow large plate-like single crystals with a typical dimension of 4 $\times$ \SI{3}{\milli \meter$^2$} in the $ab$-plane. The single crystals of EuFe$_2$As$_2$ and EuFe$_2$(As$_{1-y}$P$_y$)$_2$ were grown using a Bridgman method.\cite{Jeevan08, Jeevan10}

The structural characterization of the obtained single crystals was done by Laue diffraction and powder x-ray diffraction. The Laue method confirmed the single crystalline character of the samples. The diffraction patterns of all investigated crystals could be indexed using the ThCr$_2$Si$_2$ tetragonal type structure (space group: I4/$mmm$). Some samples exhibited a small amount of Fe$_2$As impurity phase. This amount was estimated to be less than \SI{4}{\percent}. Chemically the crystals were characterized by scanning electron microscopy equipped with energy dispersive x-ray analysis on the freshly cleaved surfaces. The inhomogeneity within the $ab$-plane was found to be around \SI{1}{\percent}. However, we found interlayer inhomogeneities along the $c$-axis which could become substantial ($\leq$ \SI{15}{\percent}). Therefore, the crystals were chosen very carefully and cleaved as thin as possible in order to reduce the error due to these interlayer inhomogeneities. Further characterization was performed by measuring the specific heat and the magnetic susceptibility of the specimens.

In general, we found that the doping of K to the Eu site turned out to be very difficult, probably due to very different chemical properties of Eu and K. The potassium tends to leave the Al$_2$O$_3$ crucible and reacts with the tantalum. Even an excess of two times the nominal value had almost no effect on the concentration of the measured crystals. Therefore, the measured doping concentrations did not exceed $x=0.52$.

Electrical resistivity, specific heat, magnetic susceptibility and thermoelectric power were measured by standard Quantum Design PPMS and MPMS systems. The contacts for the thermopower measurements were made with ordinary copper wires and a two-component epoxy, which was cured at a temperature of \SI{150}{\celsius} for \SI{10}{\minute} in an open furnace. The contact resistance was usually found to be less then \SI{1}{\ohm}. The samples were then mounted on the measurement puck in a standard four-probe configuration, as described in the Thermal Transport Option User's Manual of the PPMS.\cite{TTO_manual}


\section{Results}
\subsection{K$_x$Eu$_{1-x}$Fe$_2$As$_2$} \label{k-dope}

\begin{figure} [ht!]
\includegraphics[width= 9.4cm]{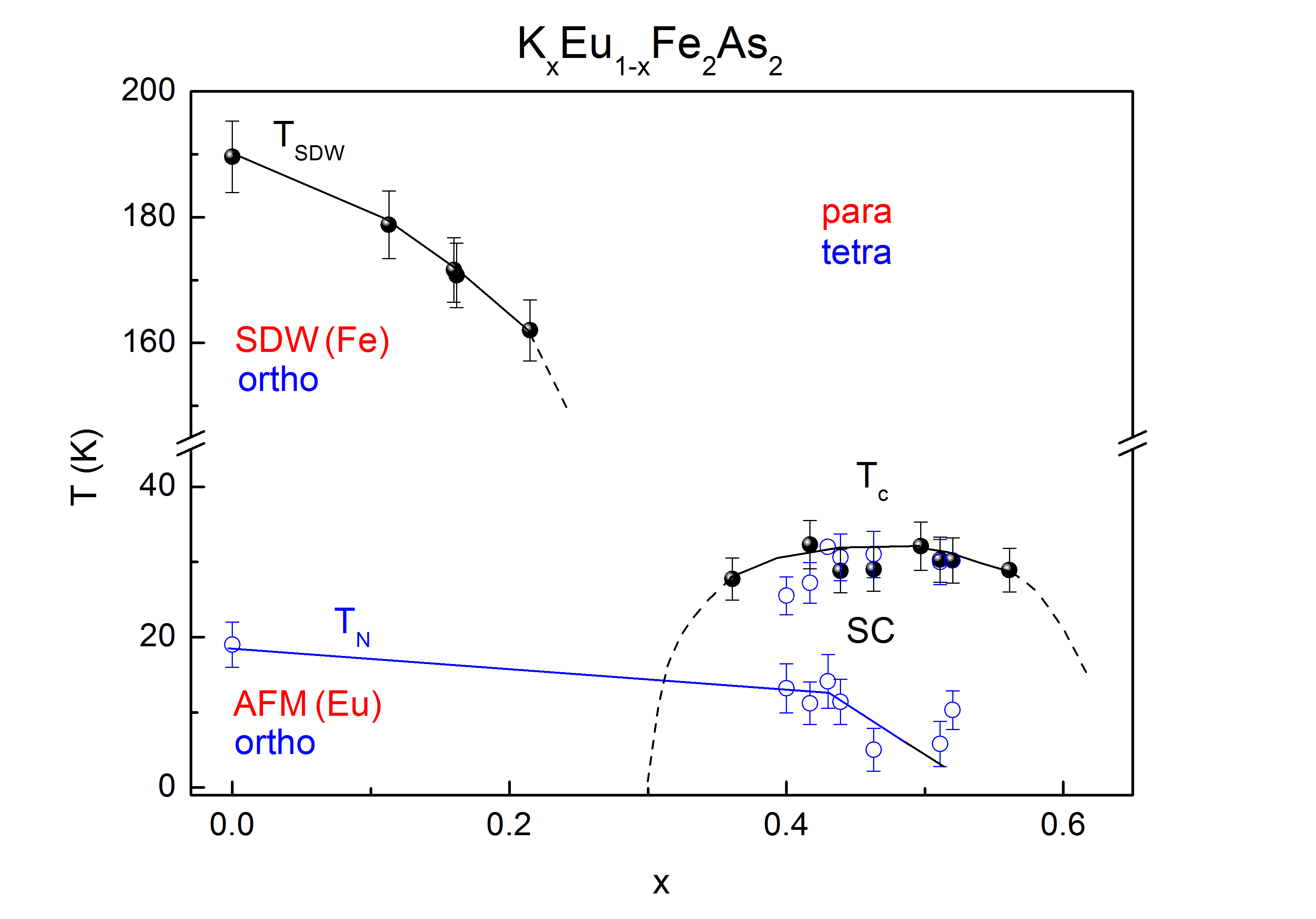}
\caption{\label{phase_diag}(Color line) The phase diagram for
K$_x$Eu$_{1-x}$Fe$_2$As$_2$ as a function of x. The black solid symbols represent phase transition temperatures obtained from resistivity measurements, while the transition temperatures marked by open blue symbols are derived from susceptibility measurements. Solid lines act as guides to the eye and dashed lines are extrapolations using the available experimental data.}
\end{figure}

\begin{figure} [ht!]
\includegraphics[width= 9.2cm]{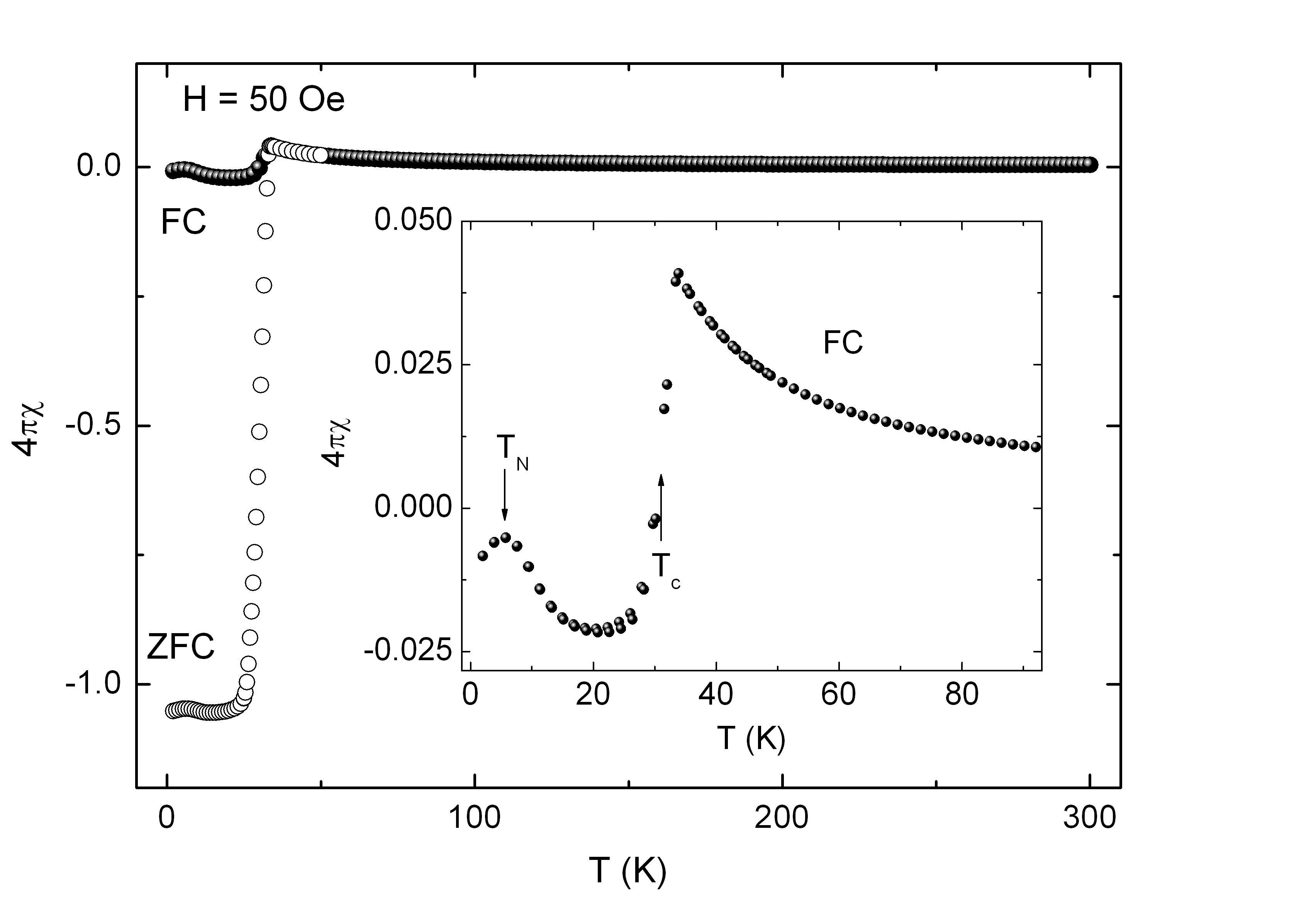}
\caption{\label{suscept} Magnetic susceptibility of K$_x$Eu$_{1-x}$Fe$_2$As$_2$ with $x=0.51$ vs. temperature. Closed and open symbols represent data obtained in a field of \SI{50}{Oe} during field cooled (FC) and zero-field cooled (ZFC) measurements. The inset enlarges the FC data
at low temperatures. Upwards and downwards pointing arrows mark $T_\mathrm{c}$ and $T_\mathrm{N}$,
respectively.}
\end{figure}

Figure~\ref{phase_diag} displays the phase diagram of K$_x$Eu$_{1-x}$Fe$_2$As$_2$ as obtained from electrical resistivity and magnetic susceptibility measurements on various
different single crystals. Both $T_{\mathrm{SDW}}$, due to Fe magnetic ordering, and $T_{\mathrm{N}}$, due to Eu magnetic ordering, lead to changes in the slope of $\rho(T)$, cf.
Fig.~\ref{double_plot_res}(a), similar as previously found on polycrystalline samples.\cite{Jeevan08b}
%
As in other iron pnictides the SDW transition is accompanied by a structural lattice distortion, which changes the crystal structure from tetragonal to orthorhombic (space group: C$mma$). We could not resolve any splitting of these transitions in our experiments. With doping, the SDW order is continuously suppressed down to approximately $T_{\mathrm{SDW}}\approx\SI{160}{\kelvin}$ where it vanishes eventually, suggesting that the SDW state might be limited to the Eu rich region of the phase diagram. For $x>0.3$, the SDW transition is completely suppressed and SC develops with a maximal $T_{\mathrm{c}}\approx \SI{34}{\kelvin}$ near $x=0.5$. Magnetic susceptibility data shown in Fig.~\ref{suscept}, provide evidence for bulk SC in this concentration regime. We note, that the broad resistive transition at x=0.22 (cf. Fig.~\ref{double_plot_res}) is probably not due to bulk SC but rather due to a minor volume fraction of higher x (see section II), which has negligible influence on the normal state properties. The data of Fig.~\ref{suscept} also indicate at least a partial AF ordering of diluted Eu$^{2+}$ moments within the SC state. Exceeding the optimal doping $T_{\mathrm{c}}$ seems to decrease again. It is probable that SC is not fully suppressed with further doping, since the end member of this doping series KFe$_2$As$_2$ shows SC, with a transition temperature of $T_{\mathrm{c}}=\SI{3.4}{\kelvin}$.\cite{Kihou10}

\subsubsection{Analysis of the electrical resistivity}
\begin{figure} [ht!]
\includegraphics[width= 9.2cm]{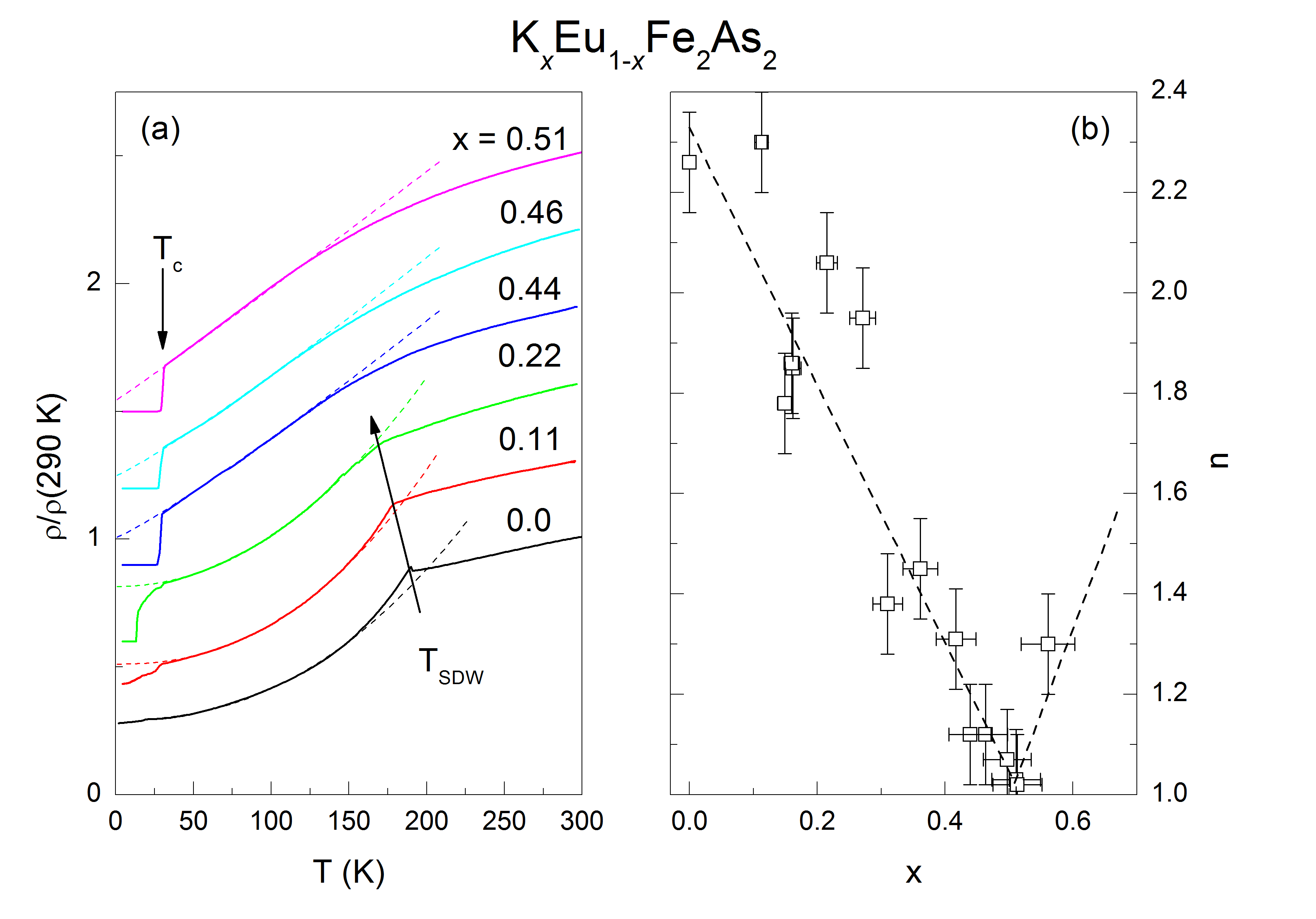}
\caption{\label{double_plot_res}(Color line) Electrical resistivity $\rho(T)/\rho(\SI{290}{\kelvin})$ vs. temperature for various different K$_x$Eu$_{1-x}$Fe$_2$As$_2$ single crystals (a). Note, that the
data are offset by $0.3$, respectively. Dashed lines indicate
$\rho(T)=\rho_0+AT^n$ dependence. The derived evolution of the exponent $n$ vs. $x$
is shown in (b). Error bars represent errors in the K-doping (see discussion
in section \ref{methods}).}
\end{figure}

The power-law behavior of the resistivity data for various compositions has been analyzed systematically, using a three-parameter fit function, $\rho(T)=\rho_0+AT^n$, to extract the effective exponent, $n$. Figure \ref{double_plot_res}(a) shows the resistivity data for some selected compositions. The validity of the fit is shown by the dashed lines. The resistivity exponent as a function of doping is shown in Fig.~\ref{double_plot_res}(b).
With increasing K-doping, $n$ decreases linearly until it reaches a minimum value of $1$ at
$x=0.51$. The exponent then starts to increases again with further doping. Such NFL behavior with exponent $n=1$ would be compatible with a two-dimensional SDW QCP. A very similar behavior has previously been found in K-doped SrFe$_2$As$_2$ polycrystals.\cite{Gooch09}

\subsubsection{Thermoelectric power}
\begin{figure} [b]
\includegraphics[width= 9.2cm]{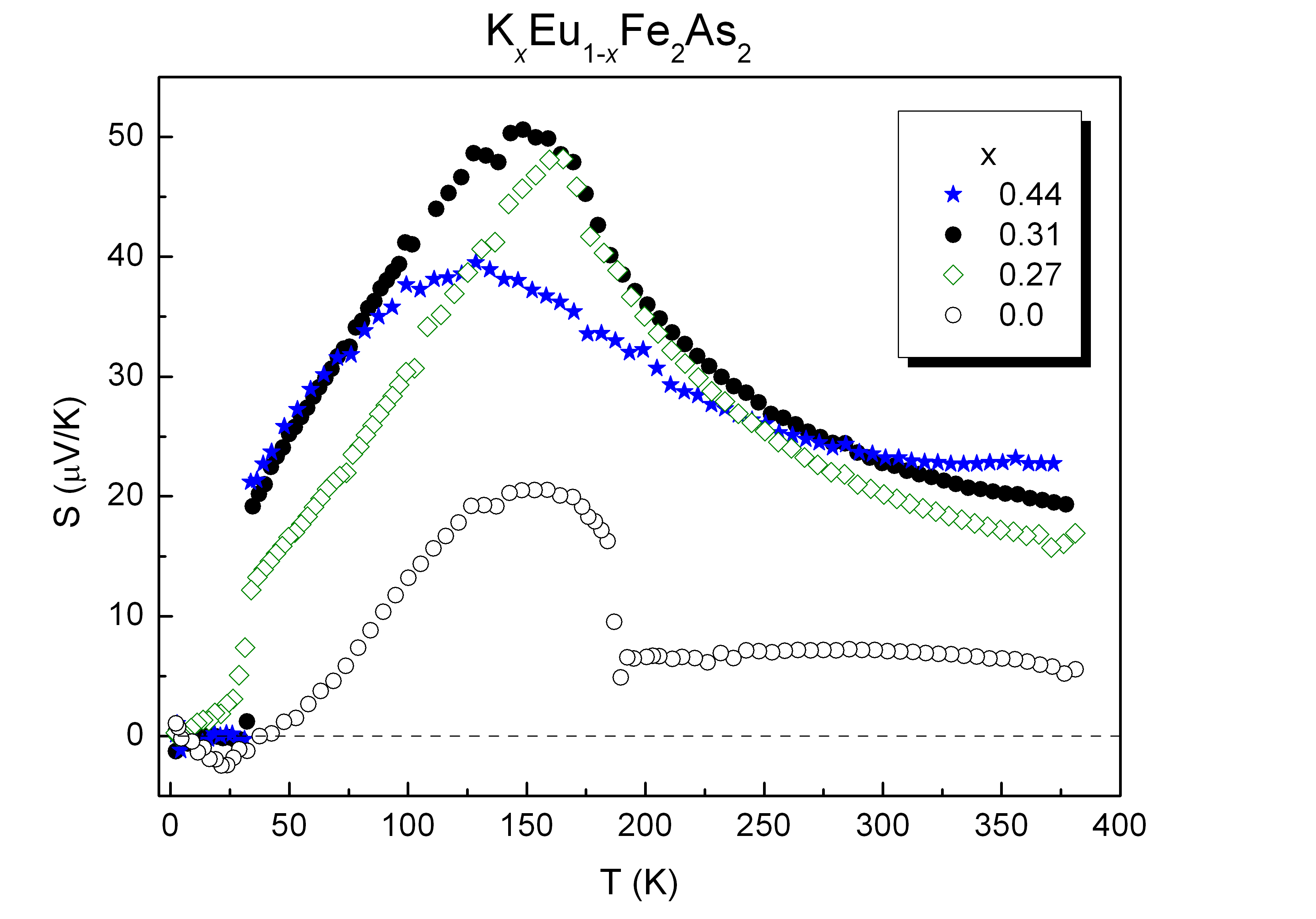}
\caption{\label{kdope_thermopower} (Color Line) Thermoelectric power $S$ vs. temperature for various different K$_x$Eu$_{1-x}$Fe$_2$As$_2$ single crystals. The dashed line marks $S=0$.}
\end{figure}
The thermoelectric power $S$ is usually the sum of the drift thermopower $S_{\mathrm{drift}}$---which is caused by charge separation due to a thermal gradient---and the phonon drag thermopower $S_{\mathrm{drag}}$---which has its origin in the electron-phonon interaction. $S_{\mathrm{drag}}$ can be an either positive or negative contribution to the total thermopower, depending on various circumstances such as fermiology, distance of the Fermi surface to the Brillouin zone boundary, dominating scattering process (normal process, umklapp process), etc.\cite{Blatt76, Barnard72}

Figure \ref{kdope_thermopower} depicts the thermopower measurements as a function of temperature for several doping compositions. At high temperatures the thermoelectric power of the parent compound EuFe$_2$As$_2$ exhibits a roughly constant positive value of approximately $\SI{7}{\micro \volt \per \kelvin}$. This might indicate that the dominant carriers are hole-like in this temperature range. However, the investigated compound is a multi-band system which consists of three hole-like Fermi surfaces at the $\bf\Gamma$-point and two electron-like Fermi surfaces at the $\bf{M}$-point. The thermopower is therefore composed of different contributions from all these bands, making a general statement difficult. Nevertheless, this interpretation seems to be supported by Hall effect measurements, which show a positive Hall coefficient above \SI{190}{\kelvin}.\cite{Ren08} Approaching the SDW transition $S(T)$ exhibits a strong increase leading to a pronounced maximum, peaked around $\SI{150}{\kelvin}$. A similar signature has previously been found e.\,g. in the Ca-122 system.\cite{Wu08} At temperatures below $\SI{50}{\kelvin}$ the thermopower crosses over to negative values and exhibits a minimum, which is centered around the N\'eel temperature of the system. However, this feature is also present in other systems (Ba-122, Sr-122, Cs-122, K-122 and even La-1111)\cite{Canfield10, Li09, Sasmal08, Lv09, Tao10} and can therefore not be related to the ordering of the Eu$^{2+}$ moments.
\begin{figure} [b]
\includegraphics[width= 9.1cm]{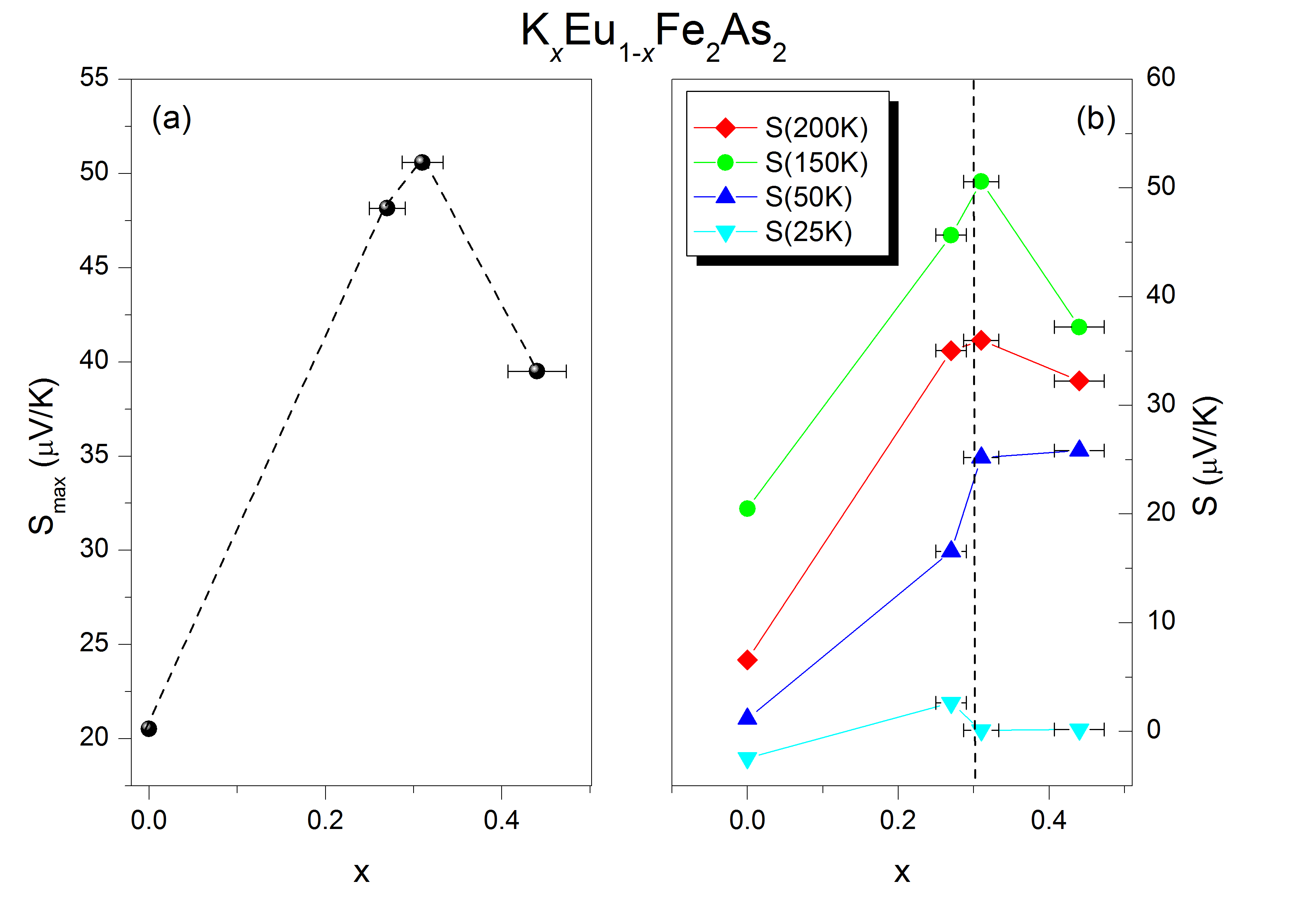}
\caption{\label{kdope_dp_S}(Color line) Maximum values of the thermoelectric power $S$ vs. K concentration derived from various different K$_x$Eu$_{1-x}$Fe$_2$As$_2$ single crystals (a).
$S$ vs. K concentration for several different temperatures is shown in (b). The dashed line in (a) acts as a guide to the eye, while in (b) it shows the critical composition $x_{\mathrm{cr}}=0.3$.}
\end{figure}
Using the reported Debye temperatures $\Theta_{\mathrm{D}}$ of the various
systems, we find that the temperature at which the minimum in $S$ is found
corresponds to $\Theta_{\mathrm{D}} / 6$. This value is well within the limits for a phonon drag peak ($\Theta_{\mathrm{D}} / 5$, $\Theta_{\mathrm{D}} / 10$).\cite{Blatt76, Barnard72} We therefore propose, that the observed feature is attributed to a negative $S_{\mathrm{drag}}$ contribution, which strongly decreases the value of $S(T)$ below $\SI{150}{\kelvin}$ and leads to the observed minimum around $\SI{20}{\kelvin}$.

The size of the thermoelectric power at room temperature increases monotonically with increasing K-doping and reaches large values of the order of \SI{50}{\micro\volt\per\kelvin}. Qualitatively, the increase of $S$ with increasing hole doping is expected in a simple rigid band picture. Such a behavior in the evolution of the band structure is already known for hole-doped Ba-122 and Sr-122.\cite{Wu11} On the other side, band structure calculations for the K$_x$Ba$_{1-x}$Fe$_2$As$_2$ system have questioned the rigid band model, indicating that the main effect of doping is a change in the relative sizes of the Fermi surfaces as a consequence of a change in As height with doping.\cite{Singh08} At the same time the peak associated with the SDW transition gets broader, confirming observations from resistivity measurements, where the SDW transition in an $\partial \rho / \partial T$ plot broadened with doping (not shown).
The maximum value of the thermopower, $S_{\mathrm{max}}(x)$, rapidly increases with hole doping and peaks at the critical composition $x_{\mathrm{cr}}=0.3$, as shown in Fig.~\ref{kdope_dp_S}(a). With further doping $S_{\mathrm{max}}$ decreases. A similar behavior has been found in K-doped Sr-122 polycrystals.\cite{Lv09b} The thermopower at fixed temperatures as a function of the control parameter, $S(x)|_{T=\mathrm{const}}$, also exhibits an anomalous behavior at the same composition, Fig.~\ref{kdope_dp_S}(b). Similar signatures have recently been observed in Co- and Ru-doped $\mathrm{BaFe_2As_2}.$\cite{Hodovanets11} The signatures at $x_\mathrm{cr}=0.3$ i.e. at the composition where
the SDW order is fully suppressed and SC emerges (cf. Fig.~\ref{phase_diag}), indicate
significant changes in the electronic structure of the system.

\begin{figure} [ht!]
\includegraphics[width= 9.2cm]{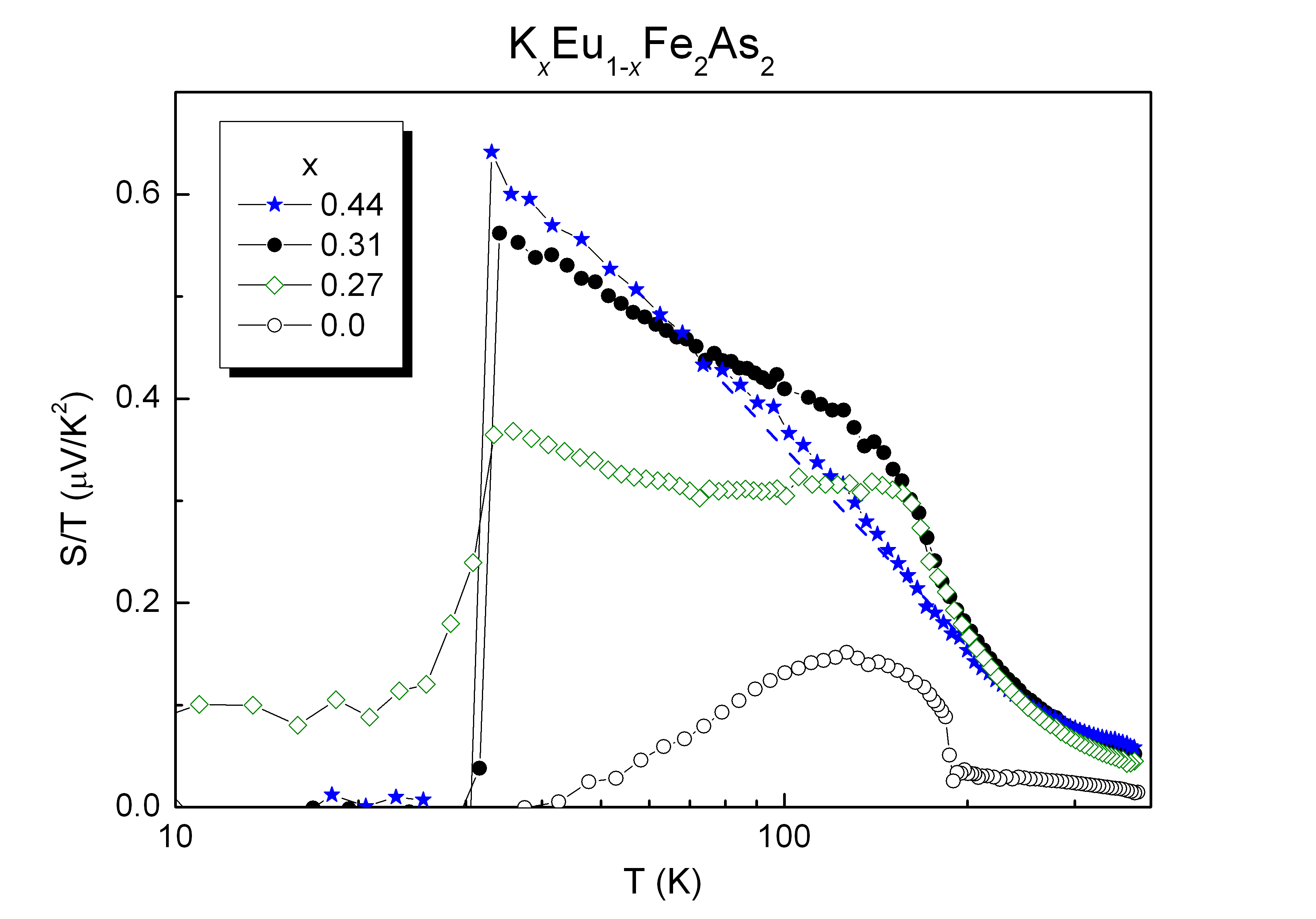}
\caption{\label{kdope_log_S}(Color line) Thermoelectric power divided by temperature $S/T$ vs. temperature (on a logarithmic scale) for various different single crystals of K$_x$Eu$_{1-x}$Fe$_2$As$_2$. The dashed line acts as a guide to the eye.}
\end{figure}

We now focus on the indications of quantum criticality in the
temperature dependence of the thermoelectric power. As shown in Fig.~\ref{kdope_log_S}, the
$x=0.44$ sample, which is located close to the optimum concentration for SC,
displays a logarithmic divergence in $S/T$ over a substantial temperature range.
Similar behavior was observed in K-doped Sr-122 polycrystals.\cite{Gooch09}

\subsection{EuFe$_2$(As$_{1-y}$P$_y$)$_2$} \label{p-dope}

\begin{figure} [ht!]
\includegraphics[width= 9.1cm]{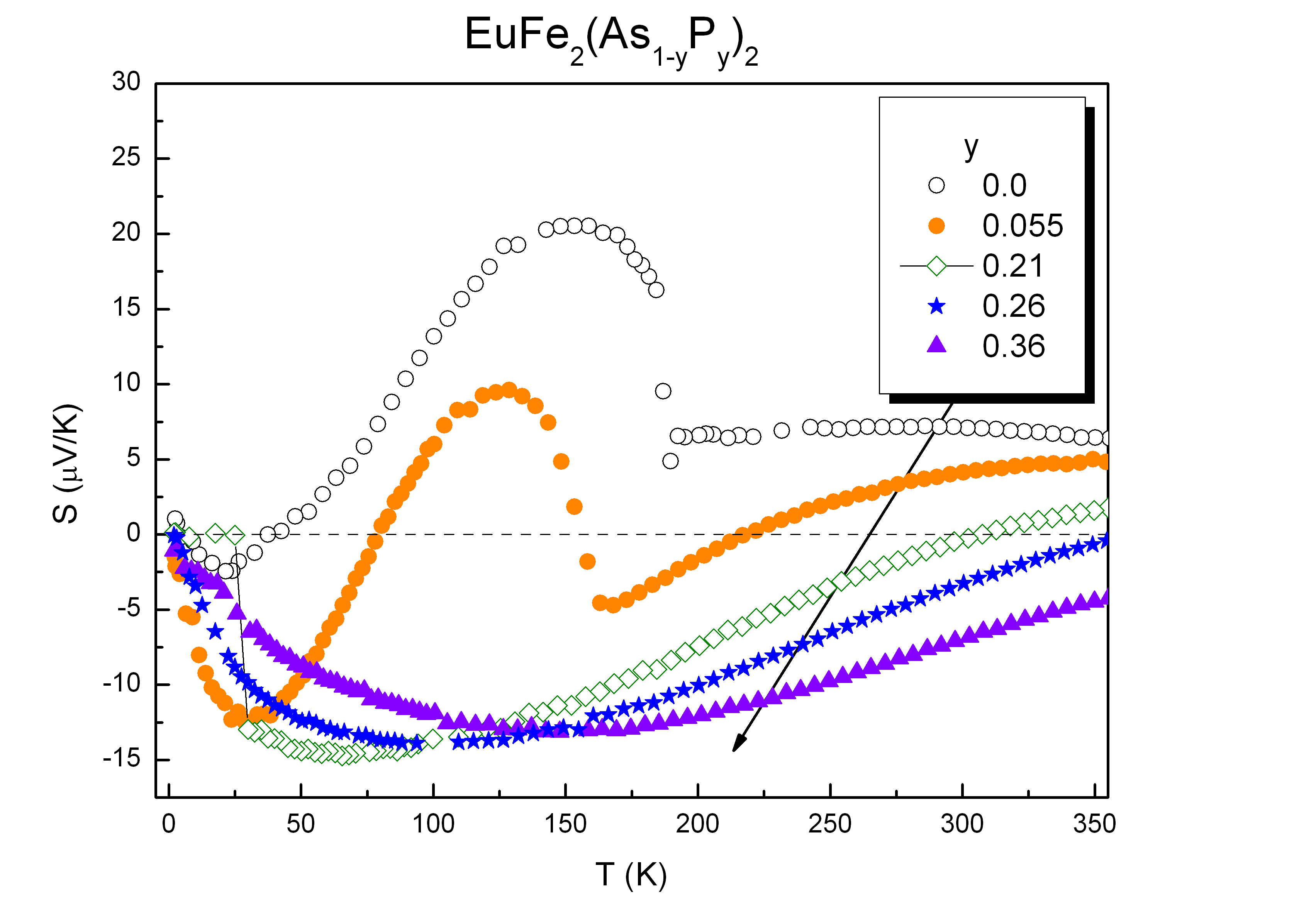}
\caption{\label{pdope_thermopower}(Color line) Thermoelectric power $S$ vs. temperature for various different EuFe$_2$(As$_{1-y}$P$_y$)$_2$ single crystals. The dashed line marks $S=0$, while the arrow indicates the increase P concentration.}
\end{figure}

The phase diagram of EuFe$_2$(As$_{1-y}$P$_y$)$_2$ has previously been determined by resistivity,
magnetic susceptibility and specific heat measurements.\cite{Jeevan10} Remarkably, SC is
confined to a very narrow regime $(0.16\le y \le 0.22)$ beyond which a ferromagnetic
Eu$^{2+}$ ordering acts detrimental. Below, we focus on thermoelectric power
measurements on similar single crystals. As shown in Fig.~\ref{pdope_thermopower}, the evolution of $S(T)$ is distinctly different to that found for K$_x$Eu$_{1-x}$Fe$_2$As$_2$.

At room temperature $S$ decreases with increasing P-substitution and changes sign at y=0.21. In an oversimplified picture this would indicate a shift from holes to electrons as the dominant carrier type. The samples  $y> 0.21$ show a negative thermopower over the whole temperature range below \SI{300}{\kelvin}, with a broad minimum at lower temperatures. The crossover from positive to negative values leads to distinct sign changes in the $y=0.055$ crystal, for which $T_\mathrm{SDW}$ is reduced to about \SI{150}{\kelvin}.

\begin{figure} 
\includegraphics[width= 9.2cm]{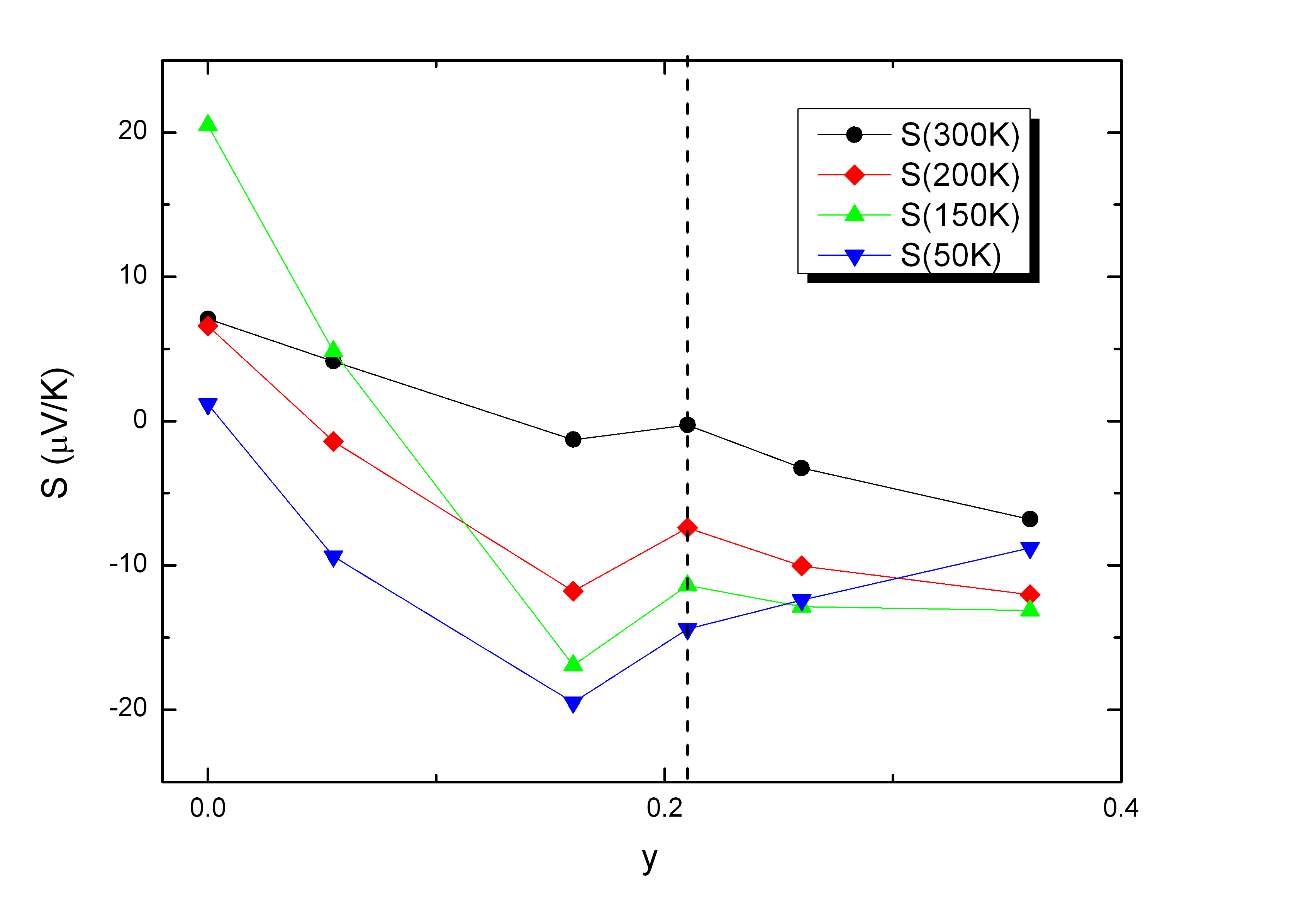}
\caption{\label{pdope_dp_S} (Color line) Thermoelectric power $S$ vs. P concentration derived from various different single crystals of EuFe$_2$(As$_{1-y}$P$_y$)$_2$. The dashed line indicates the critical composition $y_{\mathrm{cr}}=0.21$.}
\end{figure}

The evolution of the  thermopower with P-substitution, plotted in Fig.~\ref{pdope_dp_S}, indicates a distinct signature at $y_{\mathrm{cr}}=0.21$. This is close to the concentration at which SC is suppressed.\cite{Jeevan10} At this composition ARPES measurements from Thirupathaiah et al. also revealed a Lifshitz transition related to the vanishing of the inner $h$-FS at the ${\bf\Gamma}$-point.\cite{Thirupathaiah11} Recently, Liu et al. proposed that a necessary condition for SC in Co-doped Ba-122 is a nonreconstructed central hole pocket, rather than a perfect nesting condition between the central and the corner pockets.\cite{Liu11} Our data seems in agreement with this proposal.

\begin{figure} 
\includegraphics[width= 9.2cm]{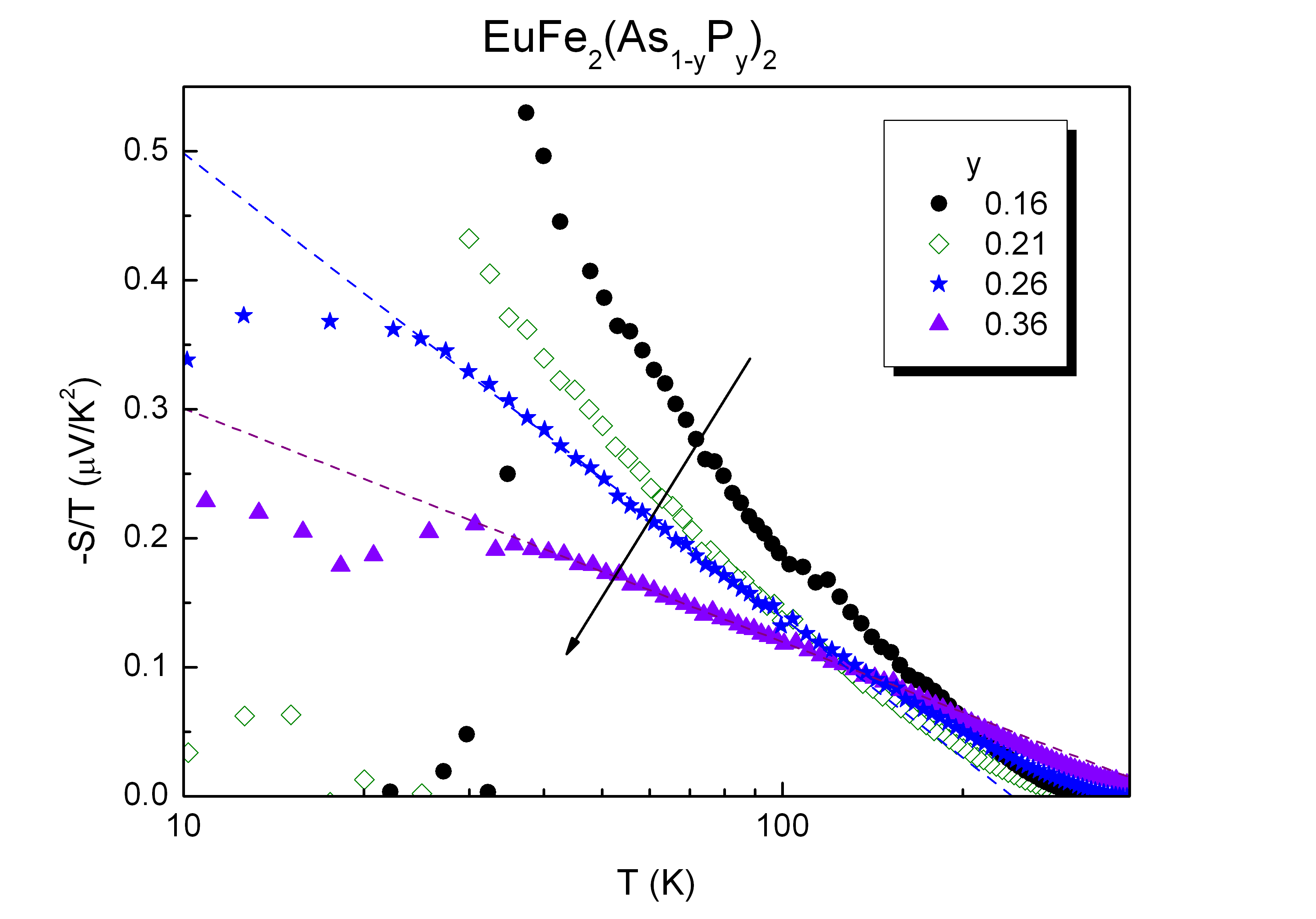}
\caption{\label{pdope_log_S} Thermoelectric power divided by temperature $S/T$ vs. temperature (on a logarithmic scale) for various different single crystals of EuFe$_2$(As$_{1-y}$P$_y$)$_2$. The dashed lines act as a guide to the eye, while the arrow indicates the increase in P concentration.}
\end{figure}

Finally, we address signatures of NFL behavior in the thermopower and electrical resistivity. Similar as in the case of the hole-doped system, the temperature dependence of $S/T$ also indicates a substantial regime (approximately one decade in $T$), for which a logarithmic divergence is observed, cf. the dashed lines in Fig.~\ref{pdope_log_S}. Most singular behavior (even weak power-law divergence) is found at low P concentrations, while with increasing y, the increase of $-S/T$ upon cooling is weaker, suggesting that the QCP is located close to  $y_{\mathrm{cr}}=0.21$.

\begin{figure} 
\includegraphics[width= 9.2cm]{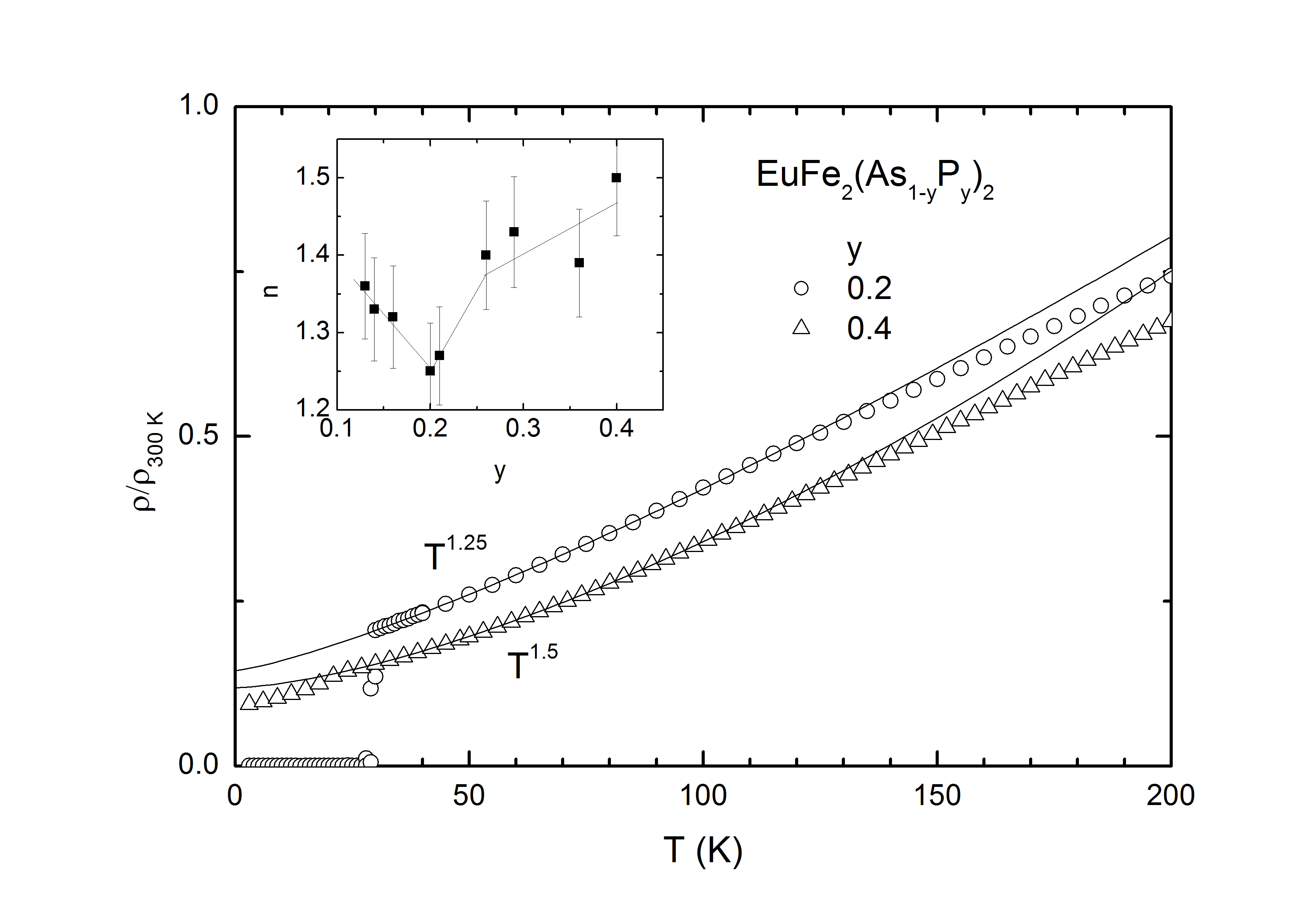}
\caption{\label{pdope_res}Electrical resistivity $\rho(T)/\rho(\SI{300}{\kelvin})$ vs. temperature for two different single crystals of EuFe$_2$(As$_{1-y}$P$_y$)$_2$ (data from  Ref.\,\onlinecite{Jeevan10}). Lines indicate $\rho(T)=\rho_0+AT^n$ dependence. The inset displays the evolution of the exponent $n(y)$ for various different  EuFe$_2$(As$_{1-y}$P$_y$)$_2$ single crystals.}
\end{figure}

Similar as for the hole-doped system, the electrical resistivity data of P-substituted single crystals from Ref.\,\onlinecite{Jeevan10} could also be described by a power-law function  $\rho(T)=\rho_0+AT^n$ at temperatures between 30 and 120~K, cf. Fig.~\ref{pdope_res}, with exponent $n<2$ indicative for NFL behavior. The evolution of $n(y)$ displays a clear minimum at $y\approx 0.2$ providing additional evidence for a QCP close to  $y_{\mathrm{cr}}$. Interestingly, the minimal value of the resistivity exponent is about 1.25 which is significantly higher than $n=1$ found for the case of hole doped  K$_{0.5}$Eu$_{0.5}$Fe$_2$As$_2$. This may suggest that the critical magnetic fluctuations, likely responsible for these NFL effects, are less two-dimensional (more isotropic) in the P-substituted case.

\section{Conclusion}

We have grown for the first time single crystals of the series K$_x$Eu$_{1-x}$Fe$_2$As$_2$ and
investigated their bulk properties with particular emphases on changes in the
electronic structure and possible indications for quantum criticality. With increasing hole doping, the thermopower at room temperature increases monotonically towards very large
values. The electrical resistivity data suggest non-Fermi liquid behavior near $x=0.5$.
This is supported by the observation of a logarithmic divergence in the
thermoelectric power coefficient $S/T$. Similar indications are also found in
single crystals of EuFe$_2$(As$_{1-y}$P$_y$)$_2$. Here, the P-substitution acts as
chemical pressure and suppresses the nesting properties of the Fermi surface
by pushing the dimensionality of the band structure towards 3D. With increasing $y$, the thermoelectric power becomes negative and displays a pronounced minimum at low temperatures. We associate this minimum, which is also present in other iron pnictide systems at the respective Debye temperatures, with a large phonon drag contribution. Furthermore, we observe a distinct signature in the thermopower evolution at $y_{\mathrm{cr}}=0.21$. A comparison
with previous ARPES measurements\cite{Thirupathaiah11} indicates that this signature is related to a Lifshitz transition, due
to the disappearance of the inner hole Fermi surface at the ${\bf\Gamma}$-point.
This change occurs at the same concentration where SC is suppressed. Therefore, our findings seem to be in agreement with the proposal by Liu et al., that a necessary condition for SC might be a nonreconstructed central hole pocket, rather than a perfect nesting between the central and the corner pockets.\cite{Liu11} In order to draw a similar conclusion for the K-doped system further investigations on its electronic structure are necessary; accordingly ARPES measurements on our single crystals are already in progress.

This work has been supported by the DFG through SPP 1458.

\bibliography{paper} 

\begin{thebibliography}{34}%
\makeatletter
\providecommand \@ifxundefined [1]{%
 \@ifx{#1\undefined}
}%
\providecommand \@ifnum [1]{%
 \ifnum #1\expandafter \@firstoftwo
 \else \expandafter \@secondoftwo
 \fi
}%
\providecommand \@ifx [1]{%
 \ifx #1\expandafter \@firstoftwo
 \else \expandafter \@secondoftwo
 \fi
}%
\providecommand \natexlab [1]{#1}%
\providecommand \enquote  [1]{``#1''}%
\providecommand \bibnamefont  [1]{#1}%
\providecommand \bibfnamefont [1]{#1}%
\providecommand \citenamefont [1]{#1}%
\providecommand \href@noop [0]{\@secondoftwo}%
\providecommand \href [0]{\begingroup \@sanitize@url \@href}%
\providecommand \@href[1]{\@@startlink{#1}\@@href}%
\providecommand \@@href[1]{\endgroup#1\@@endlink}%
\providecommand \@sanitize@url [0]{\catcode `\\12\catcode `\$12\catcode
  `\&12\catcode `\#12\catcode `\^12\catcode `\_12\catcode `\%12\relax}%
\providecommand \@@startlink[1]{}%
\providecommand \@@endlink[0]{}%
\providecommand \url  [0]{\begingroup\@sanitize@url \@url }%
\providecommand \@url [1]{\endgroup\@href {#1}{\urlprefix }}%
\providecommand \urlprefix  [0]{URL }%
\providecommand \Eprint [0]{\href }%
\providecommand \doibase [0]{http://dx.doi.org/}%
\providecommand \selectlanguage [0]{\@gobble}%
\providecommand \bibinfo  [0]{\@secondoftwo}%
\providecommand \bibfield  [0]{\@secondoftwo}%
\providecommand \translation [1]{[#1]}%
\providecommand \BibitemOpen [0]{}%
\providecommand \bibitemStop [0]{}%
\providecommand \bibitemNoStop [0]{.\EOS\space}%
\providecommand \EOS [0]{\spacefactor3000\relax}%
\providecommand \BibitemShut  [1]{\csname bibitem#1\endcsname}%
\let\auto@bib@innerbib\@empty
\bibitem [{\citenamefont {Norman}(2011)}]{Norman11}%
  \BibitemOpen
  \bibfield  {author} {\bibinfo {author} {\bibfnamefont {M.}~\bibnamefont
  {Norman}},\ }\href@noop {} {\bibfield  {journal} {\bibinfo  {journal}
  {Science}\ }\textbf {\bibinfo {volume} {332}},\ \bibinfo {pages} {196}
  (\bibinfo {year} {2011})}\BibitemShut {NoStop}%
\bibitem [{\citenamefont {Mathur}\ \emph {et~al.}(1998)\citenamefont {Mathur},
  \citenamefont {Grosche}, \citenamefont {Julian}, \citenamefont {Walker},
  \citenamefont {Freye}, \citenamefont {Haselwimmer},\ and\ \citenamefont
  {Lonzarich}}]{Mathur98}%
  \BibitemOpen
  \bibfield  {author} {\bibinfo {author} {\bibfnamefont {N.}~\bibnamefont
  {Mathur}}, \bibinfo {author} {\bibfnamefont {F.}~\bibnamefont {Grosche}},
  \bibinfo {author} {\bibfnamefont {S.}~\bibnamefont {Julian}}, \bibinfo
  {author} {\bibfnamefont {I.}~\bibnamefont {Walker}}, \bibinfo {author}
  {\bibfnamefont {D.}~\bibnamefont {Freye}}, \bibinfo {author} {\bibfnamefont
  {R.}~\bibnamefont {Haselwimmer}}, \ and\ \bibinfo {author} {\bibfnamefont
  {G.}~\bibnamefont {Lonzarich}},\ }\href@noop {} {\bibfield  {journal}
  {\bibinfo  {journal} {Nature}\ }\textbf {\bibinfo {volume} {394}},\ \bibinfo
  {pages} {39} (\bibinfo {year} {1998})}\BibitemShut {NoStop}%
\bibitem [{\citenamefont {Jeevan}\ \emph
  {et~al.}(2008{\natexlab{a}})\citenamefont {Jeevan}, \citenamefont {Hossain},
  \citenamefont {Kasinathan}, \citenamefont {Rosner}, \citenamefont {Geibel},\
  and\ \citenamefont {Gegenwart}}]{Jeevan08b}%
  \BibitemOpen
  \bibfield  {author} {\bibinfo {author} {\bibfnamefont {H.~S.}\ \bibnamefont
  {Jeevan}}, \bibinfo {author} {\bibfnamefont {Z.}~\bibnamefont {Hossain}},
  \bibinfo {author} {\bibfnamefont {D.}~\bibnamefont {Kasinathan}}, \bibinfo
  {author} {\bibfnamefont {H.}~\bibnamefont {Rosner}}, \bibinfo {author}
  {\bibfnamefont {C.}~\bibnamefont {Geibel}}, \ and\ \bibinfo {author}
  {\bibfnamefont {P.}~\bibnamefont {Gegenwart}},\ }\href@noop {} {\bibfield
  {journal} {\bibinfo  {journal} {Phys. Rev. B}\ }\textbf {\bibinfo {volume}
  {78}},\ \bibinfo {pages} {92406} (\bibinfo {year}
  {2008}{\natexlab{a}})}\BibitemShut {NoStop}%
\bibitem [{\citenamefont {Terashima}\ \emph {et~al.}(2009)\citenamefont
  {Terashima}, \citenamefont {Tomita}, \citenamefont {Kimata}, \citenamefont
  {Satsukawa}, \citenamefont {Harada}, \citenamefont {Hazama}, \citenamefont
  {Shinya}, \citenamefont {Suzuki}, \citenamefont {Matsumoto},\ and\
  \citenamefont {Murata}}]{Terashima09b}%
  \BibitemOpen
  \bibfield  {author} {\bibinfo {author} {\bibfnamefont {T.}~\bibnamefont
  {Terashima}}, \bibinfo {author} {\bibfnamefont {M.}~\bibnamefont {Tomita}},
  \bibinfo {author} {\bibfnamefont {M.}~\bibnamefont {Kimata}}, \bibinfo
  {author} {\bibfnamefont {H.}~\bibnamefont {Satsukawa}}, \bibinfo {author}
  {\bibfnamefont {A.}~\bibnamefont {Harada}}, \bibinfo {author} {\bibfnamefont
  {K.}~\bibnamefont {Hazama}}, \bibinfo {author} {\bibfnamefont
  {S.}~\bibnamefont {Shinya}, \bibfnamefont {Uji}}, \bibinfo {author}
  {\bibfnamefont {H.}~\bibnamefont {Suzuki}}, \bibinfo {author} {\bibfnamefont
  {T.}~\bibnamefont {Matsumoto}}, \ and\ \bibinfo {author} {\bibfnamefont
  {K.}~\bibnamefont {Murata}},\ }\href@noop {} {\bibfield  {journal} {\bibinfo
  {journal} {J. Phys. Soc. Jpn}\ }\textbf {\bibinfo {volume} {78}},\ \bibinfo
  {pages} {083701} (\bibinfo {year} {2009})}\BibitemShut {NoStop}%
\bibitem [{\citenamefont {Kurita}\ \emph {et~al.}(2011)\citenamefont {Kurita},
  \citenamefont {Kimata}, \citenamefont {Kodama}, \citenamefont {Harada},
  \citenamefont {Tomita}, \citenamefont {Suzuki}, \citenamefont {Matsumoto},
  \citenamefont {Murata}, \citenamefont {Uji},\ and\ \citenamefont
  {Terashima}}]{Kurita11}%
  \BibitemOpen
  \bibfield  {author} {\bibinfo {author} {\bibfnamefont {N.}~\bibnamefont
  {Kurita}}, \bibinfo {author} {\bibfnamefont {M.}~\bibnamefont {Kimata}},
  \bibinfo {author} {\bibfnamefont {K.}~\bibnamefont {Kodama}}, \bibinfo
  {author} {\bibfnamefont {A.}~\bibnamefont {Harada}}, \bibinfo {author}
  {\bibfnamefont {M.}~\bibnamefont {Tomita}}, \bibinfo {author} {\bibfnamefont
  {H.}~\bibnamefont {Suzuki}}, \bibinfo {author} {\bibfnamefont
  {T.}~\bibnamefont {Matsumoto}}, \bibinfo {author} {\bibfnamefont
  {K.}~\bibnamefont {Murata}}, \bibinfo {author} {\bibfnamefont
  {S.}~\bibnamefont {Uji}}, \ and\ \bibinfo {author} {\bibfnamefont
  {T.}~\bibnamefont {Terashima}},\ }in\ \href@noop {} {\emph {\bibinfo
  {booktitle} {Journal of Physics: Conference Series}}},\ Vol.\ \bibinfo
  {volume} {273}\ (\bibinfo {organization} {IOP Publishing},\ \bibinfo {year}
  {2011})\ p.\ \bibinfo {pages} {012098}\BibitemShut {NoStop}%
\bibitem [{\citenamefont {Qi}\ \emph {et~al.}(2008)\citenamefont {Qi},
  \citenamefont {Gao}, \citenamefont {Wang}, \citenamefont {Wang},
  \citenamefont {Zhang},\ and\ \citenamefont {Ma}}]{Qi08}%
  \BibitemOpen
  \bibfield  {author} {\bibinfo {author} {\bibfnamefont {Y.}~\bibnamefont
  {Qi}}, \bibinfo {author} {\bibfnamefont {Z.}~\bibnamefont {Gao}}, \bibinfo
  {author} {\bibfnamefont {L.}~\bibnamefont {Wang}}, \bibinfo {author}
  {\bibfnamefont {D.}~\bibnamefont {Wang}}, \bibinfo {author} {\bibfnamefont
  {X.}~\bibnamefont {Zhang}}, \ and\ \bibinfo {author} {\bibfnamefont
  {Y.}~\bibnamefont {Ma}},\ }\href@noop {} {\bibfield  {journal} {\bibinfo
  {journal} {New Journal of Physics}\ }\textbf {\bibinfo {volume} {10}},\
  \bibinfo {pages} {123003} (\bibinfo {year} {2008})}\BibitemShut {NoStop}%
\bibitem [{\citenamefont {Ren}\ \emph {et~al.}(2009)\citenamefont {Ren},
  \citenamefont {Tao}, \citenamefont {Jiang}, \citenamefont {Feng},
  \citenamefont {Wang}, \citenamefont {Dai}, \citenamefont {Cao},\ and\
  \citenamefont {Xu}}]{Ren09}%
  \BibitemOpen
  \bibfield  {author} {\bibinfo {author} {\bibfnamefont {Z.}~\bibnamefont
  {Ren}}, \bibinfo {author} {\bibfnamefont {Q.}~\bibnamefont {Tao}}, \bibinfo
  {author} {\bibfnamefont {S.}~\bibnamefont {Jiang}}, \bibinfo {author}
  {\bibfnamefont {C.}~\bibnamefont {Feng}}, \bibinfo {author} {\bibfnamefont
  {C.}~\bibnamefont {Wang}}, \bibinfo {author} {\bibfnamefont {J.}~\bibnamefont
  {Dai}}, \bibinfo {author} {\bibfnamefont {G.}~\bibnamefont {Cao}}, \ and\
  \bibinfo {author} {\bibfnamefont {Z.}~\bibnamefont {Xu}},\ }\href@noop {}
  {\bibfield  {journal} {\bibinfo  {journal} {Phys. Rev. Lett.}\ }\textbf
  {\bibinfo {volume} {102}},\ \bibinfo {pages} {137002} (\bibinfo {year}
  {2009})}\BibitemShut {NoStop}%
\bibitem [{\citenamefont {Jeevan}\ \emph
  {et~al.}(2008{\natexlab{b}})\citenamefont {Jeevan}, \citenamefont {Hossain},
  \citenamefont {Kasinathan}, \citenamefont {Rosner}, \citenamefont {Geibel},\
  and\ \citenamefont {Gegenwart}}]{Jeevan08}%
  \BibitemOpen
  \bibfield  {author} {\bibinfo {author} {\bibfnamefont {H.~S.}\ \bibnamefont
  {Jeevan}}, \bibinfo {author} {\bibfnamefont {Z.}~\bibnamefont {Hossain}},
  \bibinfo {author} {\bibfnamefont {D.}~\bibnamefont {Kasinathan}}, \bibinfo
  {author} {\bibfnamefont {H.}~\bibnamefont {Rosner}}, \bibinfo {author}
  {\bibfnamefont {C.}~\bibnamefont {Geibel}}, \ and\ \bibinfo {author}
  {\bibfnamefont {P.}~\bibnamefont {Gegenwart}},\ }\href@noop {} {\bibfield
  {journal} {\bibinfo  {journal} {Phys. Rev. B}\ }\textbf {\bibinfo {volume}
  {78}},\ \bibinfo {pages} {052502} (\bibinfo {year}
  {2008}{\natexlab{b}})}\BibitemShut {NoStop}%
\bibitem [{\citenamefont {Herrero-Mart{\'\i}n}\ \emph
  {et~al.}(2009)\citenamefont {Herrero-Mart{\'\i}n}, \citenamefont {Scagnoli},
  \citenamefont {Mazzoli}, \citenamefont {Su}, \citenamefont {Mittal},
  \citenamefont {Xiao}, \citenamefont {Brueckel}, \citenamefont {Kumar},
  \citenamefont {Dhar}, \citenamefont {Thamizhavel},\ and\ \citenamefont
  {Paolasini}}]{Herrero09}%
  \BibitemOpen
  \bibfield  {author} {\bibinfo {author} {\bibfnamefont {J.}~\bibnamefont
  {Herrero-Mart{\'\i}n}}, \bibinfo {author} {\bibfnamefont {V.}~\bibnamefont
  {Scagnoli}}, \bibinfo {author} {\bibfnamefont {C.}~\bibnamefont {Mazzoli}},
  \bibinfo {author} {\bibfnamefont {Y.}~\bibnamefont {Su}}, \bibinfo {author}
  {\bibfnamefont {R.}~\bibnamefont {Mittal}}, \bibinfo {author} {\bibfnamefont
  {Y.}~\bibnamefont {Xiao}}, \bibinfo {author} {\bibfnamefont {T.}~\bibnamefont
  {Brueckel}}, \bibinfo {author} {\bibfnamefont {N.}~\bibnamefont {Kumar}},
  \bibinfo {author} {\bibfnamefont {S.~K.}\ \bibnamefont {Dhar}}, \bibinfo
  {author} {\bibfnamefont {A.}~\bibnamefont {Thamizhavel}}, \ and\ \bibinfo
  {author} {\bibfnamefont {L.}~\bibnamefont {Paolasini}},\ }\href@noop {}
  {\bibfield  {journal} {\bibinfo  {journal} {Phys. Rev. B}\ }\textbf {\bibinfo
  {volume} {80}},\ \bibinfo {pages} {134411} (\bibinfo {year}
  {2009})}\BibitemShut {NoStop}%
\bibitem [{\citenamefont {Wen}(2011)}]{Wen11}%
  \BibitemOpen
  \bibfield  {author} {\bibinfo {author} {\bibfnamefont {H.}~\bibnamefont
  {Wen}},\ }\href@noop {} {\bibfield  {journal} {\bibinfo  {journal} {Annual
  Review of Condensed Matter Physics}\ }\textbf {\bibinfo {volume} {2}}
  (\bibinfo {year} {2011})}\BibitemShut {NoStop}%
\bibitem [{\citenamefont {Johnston}(2010)}]{Johnston10}%
  \BibitemOpen
  \bibfield  {author} {\bibinfo {author} {\bibfnamefont {D.}~\bibnamefont
  {Johnston}},\ }\href@noop {} {\bibfield  {journal} {\bibinfo  {journal}
  {Advances in Physics}\ }\textbf {\bibinfo {volume} {59}},\ \bibinfo {pages}
  {803} (\bibinfo {year} {2010})}\BibitemShut {NoStop}%
\bibitem [{\citenamefont {Paglione}\ and\ \citenamefont
  {Greene}(2010)}]{Paglione10}%
  \BibitemOpen
  \bibfield  {author} {\bibinfo {author} {\bibfnamefont {J.}~\bibnamefont
  {Paglione}}\ and\ \bibinfo {author} {\bibfnamefont {R.}~\bibnamefont
  {Greene}},\ }\href@noop {} {\bibfield  {journal} {\bibinfo  {journal} {Nature
  Physics}\ }\textbf {\bibinfo {volume} {6}},\ \bibinfo {pages} {645} (\bibinfo
  {year} {2010})}\BibitemShut {NoStop}%
\bibitem [{\citenamefont {Jeevan}\ \emph {et~al.}(2011)\citenamefont {Jeevan},
  \citenamefont {Kasinathan}, \citenamefont {Rosner},\ and\ \citenamefont
  {Gegenwart}}]{Jeevan10}%
  \BibitemOpen
  \bibfield  {author} {\bibinfo {author} {\bibfnamefont {H.~S.}\ \bibnamefont
  {Jeevan}}, \bibinfo {author} {\bibfnamefont {D.}~\bibnamefont {Kasinathan}},
  \bibinfo {author} {\bibfnamefont {H.}~\bibnamefont {Rosner}}, \ and\ \bibinfo
  {author} {\bibfnamefont {P.}~\bibnamefont {Gegenwart}},\ }\href@noop {}
  {\bibfield  {journal} {\bibinfo  {journal} {Phys. Rev. B}\ }\textbf {\bibinfo
  {volume} {83}},\ \bibinfo {pages} {054511} (\bibinfo {year}
  {2011})}\BibitemShut {NoStop}%
\bibitem [{\citenamefont {Thirupathaiah}\ \emph {et~al.}(2011)\citenamefont
  {Thirupathaiah}, \citenamefont {Rienks}, \citenamefont {Jeevan},
  \citenamefont {Ovsyannikov}, \citenamefont {Slooten}, \citenamefont {Kaas},
  \citenamefont {van Heumen}, \citenamefont {de~Jong}, \citenamefont {D\"urr},
  \citenamefont {Siemensmeyer}, \citenamefont {Follath}, \citenamefont
  {Gegenwart}, \citenamefont {Golden},\ and\ \citenamefont
  {Fink}}]{Thirupathaiah11}%
  \BibitemOpen
  \bibfield  {author} {\bibinfo {author} {\bibfnamefont {S.}~\bibnamefont
  {Thirupathaiah}}, \bibinfo {author} {\bibfnamefont {E.~D.~L.}\ \bibnamefont
  {Rienks}}, \bibinfo {author} {\bibfnamefont {H.~S.}\ \bibnamefont {Jeevan}},
  \bibinfo {author} {\bibfnamefont {R.}~\bibnamefont {Ovsyannikov}}, \bibinfo
  {author} {\bibfnamefont {E.}~\bibnamefont {Slooten}}, \bibinfo {author}
  {\bibfnamefont {J.}~\bibnamefont {Kaas}}, \bibinfo {author} {\bibfnamefont
  {E.}~\bibnamefont {van Heumen}}, \bibinfo {author} {\bibfnamefont
  {S.}~\bibnamefont {de~Jong}}, \bibinfo {author} {\bibfnamefont {H.~A.}\
  \bibnamefont {D\"urr}}, \bibinfo {author} {\bibfnamefont {K.}~\bibnamefont
  {Siemensmeyer}}, \bibinfo {author} {\bibfnamefont {R.}~\bibnamefont
  {Follath}}, \bibinfo {author} {\bibfnamefont {P.}~\bibnamefont {Gegenwart}},
  \bibinfo {author} {\bibfnamefont {M.~S.}\ \bibnamefont {Golden}}, \ and\
  \bibinfo {author} {\bibfnamefont {J.}~\bibnamefont {Fink}},\ }\href {\doibase
  10.1103/PhysRevB.84.014531} {\bibfield  {journal} {\bibinfo  {journal} {Phys.
  Rev. B}\ }\textbf {\bibinfo {volume} {84}},\ \bibinfo {pages} {014531}
  (\bibinfo {year} {2011})}\BibitemShut {NoStop}%
\bibitem [{\citenamefont {Wu}\ \emph {et~al.}(2011)\citenamefont {Wu},
  \citenamefont {Chanda}, \citenamefont {Jeevan}, \citenamefont {Gegenwart},\
  and\ \citenamefont {Dressel}}]{Wu11}%
  \BibitemOpen
  \bibfield  {author} {\bibinfo {author} {\bibfnamefont {D.}~\bibnamefont
  {Wu}}, \bibinfo {author} {\bibfnamefont {G.}~\bibnamefont {Chanda}}, \bibinfo
  {author} {\bibfnamefont {H.~S.}\ \bibnamefont {Jeevan}}, \bibinfo {author}
  {\bibfnamefont {P.}~\bibnamefont {Gegenwart}}, \ and\ \bibinfo {author}
  {\bibfnamefont {M.}~\bibnamefont {Dressel}},\ }\href@noop {} {\bibfield
  {journal} {\bibinfo  {journal} {Phys. Rev. B}\ }\textbf {\bibinfo {volume}
  {83}},\ \bibinfo {pages} {100503} (\bibinfo {year} {2011})}\BibitemShut
  {NoStop}%
\bibitem [{\citenamefont {Paul}\ and\ \citenamefont {Kotliar}(2001)}]{Paul01}%
  \BibitemOpen
  \bibfield  {author} {\bibinfo {author} {\bibfnamefont {I.}~\bibnamefont
  {Paul}}\ and\ \bibinfo {author} {\bibfnamefont {G.}~\bibnamefont {Kotliar}},\
  }\href@noop {} {\bibfield  {journal} {\bibinfo  {journal} {Phys. Rev. B}\
  }\textbf {\bibinfo {volume} {64}},\ \bibinfo {pages} {184414} (\bibinfo
  {year} {2001})}\BibitemShut {NoStop}%
\bibitem [{\citenamefont {Kondo}(2002)}]{Kondo02}%
  \BibitemOpen
  \bibfield  {author} {\bibinfo {author} {\bibfnamefont {H.}~\bibnamefont
  {Kondo}},\ }\href@noop {} {\bibfield  {journal} {\bibinfo  {journal} {J.
  Phys. Soc. Jpn.}\ }\textbf {\bibinfo {volume} {71}},\ \bibinfo {pages} {3011}
  (\bibinfo {year} {2002})},\ \bibinfo {note} {and references
  therein}\BibitemShut {NoStop}%
\bibitem [{\citenamefont {Zapf}\ \emph {et~al.}(2011)\citenamefont {Zapf},
  \citenamefont {Wu}, \citenamefont {Bogani}, \citenamefont {Jeevan},
  \citenamefont {Gegenwart},\ and\ \citenamefont {Dressel}}]{Zapf11}%
  \BibitemOpen
  \bibfield  {author} {\bibinfo {author} {\bibfnamefont {S.}~\bibnamefont
  {Zapf}}, \bibinfo {author} {\bibfnamefont {D.}~\bibnamefont {Wu}}, \bibinfo
  {author} {\bibfnamefont {L.}~\bibnamefont {Bogani}}, \bibinfo {author}
  {\bibfnamefont {H.~S.}\ \bibnamefont {Jeevan}}, \bibinfo {author}
  {\bibfnamefont {P.}~\bibnamefont {Gegenwart}}, \ and\ \bibinfo {author}
  {\bibfnamefont {M.}~\bibnamefont {Dressel}},\ }\href@noop {} {\bibfield
  {journal} {\bibinfo  {journal} {Phys. Rev. B}\ }\textbf {\bibinfo {volume}
  {84}},\ \bibinfo {pages} {140503} (\bibinfo {year} {2011})}\BibitemShut
  {NoStop}%
\bibitem [{TTO(2002)}]{TTO_manual}%
  \BibitemOpen
  \href@noop {} {\emph {\bibinfo {title} {Physical Property Measurement System:
  Thermal Transport Option User's Manual}}},\ \bibinfo {organization} {Quantum
  Design},\ \bibinfo {address} {11578 Sorrento Valley Rd. San Diego, CA
  92121-1311 USA},\ \bibinfo {edition} {3rd}\ ed. (\bibinfo {year}
  {2002})\BibitemShut {NoStop}%
\bibitem [{\citenamefont {Kihou}\ \emph {et~al.}(2010)\citenamefont {Kihou},
  \citenamefont {Saito}, \citenamefont {Ishida}, \citenamefont {Nakajima},
  \citenamefont {Tomioka}, \citenamefont {Fukazawa}, \citenamefont {Kohori},
  \citenamefont {Ito}, \citenamefont {Uchida}, \citenamefont {Iyo} \emph
  {et~al.}}]{Kihou10}%
  \BibitemOpen
  \bibfield  {author} {\bibinfo {author} {\bibfnamefont {K.}~\bibnamefont
  {Kihou}}, \bibinfo {author} {\bibfnamefont {T.}~\bibnamefont {Saito}},
  \bibinfo {author} {\bibfnamefont {S.}~\bibnamefont {Ishida}}, \bibinfo
  {author} {\bibfnamefont {M.}~\bibnamefont {Nakajima}}, \bibinfo {author}
  {\bibfnamefont {Y.}~\bibnamefont {Tomioka}}, \bibinfo {author} {\bibfnamefont
  {H.}~\bibnamefont {Fukazawa}}, \bibinfo {author} {\bibfnamefont
  {Y.}~\bibnamefont {Kohori}}, \bibinfo {author} {\bibfnamefont
  {T.}~\bibnamefont {Ito}}, \bibinfo {author} {\bibfnamefont {S.}~\bibnamefont
  {Uchida}}, \bibinfo {author} {\bibfnamefont {A.}~\bibnamefont {Iyo}},  \emph
  {et~al.},\ }\href@noop {} {\bibfield  {journal} {\bibinfo  {journal} {J.
  Phys. Soc. Jpn.}\ }\textbf {\bibinfo {volume} {79}},\ \bibinfo {pages}
  {124713} (\bibinfo {year} {2010})}\BibitemShut {NoStop}%
\bibitem [{\citenamefont {Gooch}\ \emph {et~al.}(2009)\citenamefont {Gooch},
  \citenamefont {Lv}, \citenamefont {Lorenz}, \citenamefont {Guloy},\ and\
  \citenamefont {Chu}}]{Gooch09}%
  \BibitemOpen
  \bibfield  {author} {\bibinfo {author} {\bibfnamefont {M.}~\bibnamefont
  {Gooch}}, \bibinfo {author} {\bibfnamefont {B.}~\bibnamefont {Lv}}, \bibinfo
  {author} {\bibfnamefont {B.}~\bibnamefont {Lorenz}}, \bibinfo {author}
  {\bibfnamefont {A.~M.}\ \bibnamefont {Guloy}}, \ and\ \bibinfo {author}
  {\bibfnamefont {C.~W.}\ \bibnamefont {Chu}},\ }\href@noop {} {\bibfield
  {journal} {\bibinfo  {journal} {Phys. Rev. B}\ }\textbf {\bibinfo {volume}
  {79}},\ \bibinfo {pages} {104504} (\bibinfo {year} {2009})}\BibitemShut
  {NoStop}%
\bibitem [{\citenamefont {Blatt}\ \emph {et~al.}(1976)\citenamefont {Blatt},
  \citenamefont {Schroeder}, \citenamefont {Foiles},\ and\ \citenamefont
  {Greig}}]{Blatt76}%
  \BibitemOpen
  \bibfield  {author} {\bibinfo {author} {\bibfnamefont {F.~J.}\ \bibnamefont
  {Blatt}}, \bibinfo {author} {\bibfnamefont {P.~A.}\ \bibnamefont
  {Schroeder}}, \bibinfo {author} {\bibfnamefont {C.~L.}\ \bibnamefont
  {Foiles}}, \ and\ \bibinfo {author} {\bibfnamefont {D.}~\bibnamefont
  {Greig}},\ }\href@noop {} {\emph {\bibinfo {title} {Thermoelectric Power of
  Metals}}}\ (\bibinfo  {publisher} {Plenum Press, New York},\ \bibinfo {year}
  {1976})\BibitemShut {NoStop}%
\bibitem [{\citenamefont {Barnard}(1972)}]{Barnard72}%
  \BibitemOpen
  \bibfield  {author} {\bibinfo {author} {\bibfnamefont {R.~D.}\ \bibnamefont
  {Barnard}},\ }\href@noop {} {\emph {\bibinfo {title} {Thermoelectricity in
  Metals and Alloys}}}\ (\bibinfo  {publisher} {Taylor \& Francis Ltd.,
  London},\ \bibinfo {year} {1972})\BibitemShut {NoStop}%
\bibitem [{\citenamefont {Ren}\ \emph {et~al.}(2008)\citenamefont {Ren},
  \citenamefont {Zhu}, \citenamefont {Jiang}, \citenamefont {Xu}, \citenamefont
  {Tao}, \citenamefont {Wang}, \citenamefont {Feng}, \citenamefont {Cao},\ and\
  \citenamefont {Xu}}]{Ren08}%
  \BibitemOpen
  \bibfield  {author} {\bibinfo {author} {\bibfnamefont {Z.}~\bibnamefont
  {Ren}}, \bibinfo {author} {\bibfnamefont {Z.}~\bibnamefont {Zhu}}, \bibinfo
  {author} {\bibfnamefont {S.}~\bibnamefont {Jiang}}, \bibinfo {author}
  {\bibfnamefont {X.}~\bibnamefont {Xu}}, \bibinfo {author} {\bibfnamefont
  {Q.}~\bibnamefont {Tao}}, \bibinfo {author} {\bibfnamefont {C.}~\bibnamefont
  {Wang}}, \bibinfo {author} {\bibfnamefont {C.}~\bibnamefont {Feng}}, \bibinfo
  {author} {\bibfnamefont {G.}~\bibnamefont {Cao}}, \ and\ \bibinfo {author}
  {\bibfnamefont {Z.}~\bibnamefont {Xu}},\ }\href@noop {} {\bibfield  {journal}
  {\bibinfo  {journal} {Phys. Rev. B}\ }\textbf {\bibinfo {volume} {78}},\
  \bibinfo {pages} {052501} (\bibinfo {year} {2008})}\BibitemShut {NoStop}%
\bibitem [{\citenamefont {Wu}\ \emph {et~al.}(2008)\citenamefont {Wu},
  \citenamefont {Chen}, \citenamefont {Wu}, \citenamefont {Xie}, \citenamefont
  {Yan}, \citenamefont {Liu}, \citenamefont {Wang}, \citenamefont {Ying},\ and\
  \citenamefont {Chen}}]{Wu08}%
  \BibitemOpen
  \bibfield  {author} {\bibinfo {author} {\bibfnamefont {G.}~\bibnamefont
  {Wu}}, \bibinfo {author} {\bibfnamefont {H.}~\bibnamefont {Chen}}, \bibinfo
  {author} {\bibfnamefont {T.}~\bibnamefont {Wu}}, \bibinfo {author}
  {\bibfnamefont {Y.}~\bibnamefont {Xie}}, \bibinfo {author} {\bibfnamefont
  {Y.}~\bibnamefont {Yan}}, \bibinfo {author} {\bibfnamefont {R.}~\bibnamefont
  {Liu}}, \bibinfo {author} {\bibfnamefont {X.}~\bibnamefont {Wang}}, \bibinfo
  {author} {\bibfnamefont {J.}~\bibnamefont {Ying}}, \ and\ \bibinfo {author}
  {\bibfnamefont {X.}~\bibnamefont {Chen}},\ }\href@noop {} {\bibfield
  {journal} {\bibinfo  {journal} {J. Phys.: Condens. Matter}\ }\textbf
  {\bibinfo {volume} {20}},\ \bibinfo {pages} {422201} (\bibinfo {year}
  {2008})}\BibitemShut {NoStop}%
\bibitem [{\citenamefont {Canfield}\ and\ \citenamefont
  {Bud'ko}(2010)}]{Canfield10}%
  \BibitemOpen
  \bibfield  {author} {\bibinfo {author} {\bibfnamefont {P.~C.}\ \bibnamefont
  {Canfield}}\ and\ \bibinfo {author} {\bibfnamefont {S.~L.}\ \bibnamefont
  {Bud'ko}},\ }\href {\doibase 10.1146/annurev-conmatphys-070909-104041}
  {\bibfield  {journal} {\bibinfo  {journal} {Ann. Rev. Condens. Mat. Phys.}\
  }\textbf {\bibinfo {volume} {1}},\ \bibinfo {pages} {27} (\bibinfo {year}
  {2010})}\BibitemShut {NoStop}%
\bibitem [{\citenamefont {Li}\ \emph {et~al.}(2009)\citenamefont {Li},
  \citenamefont {Luo}, \citenamefont {Wang}, \citenamefont {Chen},
  \citenamefont {Ren}, \citenamefont {Tao}, \citenamefont {Li}, \citenamefont
  {Lin}, \citenamefont {He}, \citenamefont {Zhu} \emph {et~al.}}]{Li09}%
  \BibitemOpen
  \bibfield  {author} {\bibinfo {author} {\bibfnamefont {L.}~\bibnamefont
  {Li}}, \bibinfo {author} {\bibfnamefont {Y.}~\bibnamefont {Luo}}, \bibinfo
  {author} {\bibfnamefont {Q.}~\bibnamefont {Wang}}, \bibinfo {author}
  {\bibfnamefont {H.}~\bibnamefont {Chen}}, \bibinfo {author} {\bibfnamefont
  {Z.}~\bibnamefont {Ren}}, \bibinfo {author} {\bibfnamefont {Q.}~\bibnamefont
  {Tao}}, \bibinfo {author} {\bibfnamefont {Y.}~\bibnamefont {Li}}, \bibinfo
  {author} {\bibfnamefont {X.}~\bibnamefont {Lin}}, \bibinfo {author}
  {\bibfnamefont {M.}~\bibnamefont {He}}, \bibinfo {author} {\bibfnamefont
  {Z.}~\bibnamefont {Zhu}},  \emph {et~al.},\ }\href@noop {} {\bibfield
  {journal} {\bibinfo  {journal} {New Journal of Physics}\ }\textbf {\bibinfo
  {volume} {11}},\ \bibinfo {pages} {025008} (\bibinfo {year}
  {2009})}\BibitemShut {NoStop}%
\bibitem [{\citenamefont {Sasmal}\ \emph {et~al.}(2008)\citenamefont {Sasmal},
  \citenamefont {Lv}, \citenamefont {Lorenz}, \citenamefont {Guloy},
  \citenamefont {Chen}, \citenamefont {Xue},\ and\ \citenamefont
  {Chu}}]{Sasmal08}%
  \BibitemOpen
  \bibfield  {author} {\bibinfo {author} {\bibfnamefont {K.}~\bibnamefont
  {Sasmal}}, \bibinfo {author} {\bibfnamefont {B.}~\bibnamefont {Lv}}, \bibinfo
  {author} {\bibfnamefont {B.}~\bibnamefont {Lorenz}}, \bibinfo {author}
  {\bibfnamefont {A.~M.}\ \bibnamefont {Guloy}}, \bibinfo {author}
  {\bibfnamefont {F.}~\bibnamefont {Chen}}, \bibinfo {author} {\bibfnamefont
  {Y.-Y.}\ \bibnamefont {Xue}}, \ and\ \bibinfo {author} {\bibfnamefont
  {C.-W.}\ \bibnamefont {Chu}},\ }\href {\doibase
  10.1103/PhysRevLett.101.107007} {\bibfield  {journal} {\bibinfo  {journal}
  {Phys. Rev. Lett.}\ }\textbf {\bibinfo {volume} {101}},\ \bibinfo {pages}
  {107007} (\bibinfo {year} {2008})}\BibitemShut {NoStop}%
\bibitem [{\citenamefont {Lv}(2009)}]{Lv09}%
  \BibitemOpen
  \bibfield  {author} {\bibinfo {author} {\bibfnamefont {B.}~\bibnamefont
  {Lv}},\ }\emph {\bibinfo {title} {Syntheses, Structures and Properties of
  FeAs-based Superconductors}},\ \href@noop {} {Ph.D. thesis},\ \bibinfo
  {school} {University of Houston´} (\bibinfo {year} {2009})\BibitemShut
  {NoStop}%
\bibitem [{\citenamefont {Tao}\ \emph {et~al.}(2010)\citenamefont {Tao},
  \citenamefont {Zhu}, \citenamefont {Lin}, \citenamefont {Cao}, \citenamefont
  {Xu}, \citenamefont {Chen}, \citenamefont {Luo},\ and\ \citenamefont
  {Wang}}]{Tao10}%
  \BibitemOpen
  \bibfield  {author} {\bibinfo {author} {\bibfnamefont {Q.}~\bibnamefont
  {Tao}}, \bibinfo {author} {\bibfnamefont {Z.}~\bibnamefont {Zhu}}, \bibinfo
  {author} {\bibfnamefont {X.}~\bibnamefont {Lin}}, \bibinfo {author}
  {\bibfnamefont {G.}~\bibnamefont {Cao}}, \bibinfo {author} {\bibfnamefont
  {Z.}~\bibnamefont {Xu}}, \bibinfo {author} {\bibfnamefont {G.}~\bibnamefont
  {Chen}}, \bibinfo {author} {\bibfnamefont {J.}~\bibnamefont {Luo}}, \ and\
  \bibinfo {author} {\bibfnamefont {N.}~\bibnamefont {Wang}},\ }\href@noop {}
  {\bibfield  {journal} {\bibinfo  {journal} {J. Phys.: Condens. Mattter}\
  }\textbf {\bibinfo {volume} {22}},\ \bibinfo {pages} {072201} (\bibinfo
  {year} {2010})}\BibitemShut {NoStop}%
\bibitem [{\citenamefont {Singh}(2008)}]{Singh08}%
  \BibitemOpen
  \bibfield  {author} {\bibinfo {author} {\bibfnamefont {D.~J.}\ \bibnamefont
  {Singh}},\ }\href@noop {} {\bibfield  {journal} {\bibinfo  {journal} {Phys.
  Rev. B}\ }\textbf {\bibinfo {volume} {78}},\ \bibinfo {pages} {94511}
  (\bibinfo {year} {2008})}\BibitemShut {NoStop}%
\bibitem [{\citenamefont {Lv}\ \emph {et~al.}(2009)\citenamefont {Lv},
  \citenamefont {Gooch}, \citenamefont {Lorenz}, \citenamefont {Chen},
  \citenamefont {Guloy},\ and\ \citenamefont {Chu}}]{Lv09b}%
  \BibitemOpen
  \bibfield  {author} {\bibinfo {author} {\bibfnamefont {B.}~\bibnamefont
  {Lv}}, \bibinfo {author} {\bibfnamefont {M.}~\bibnamefont {Gooch}}, \bibinfo
  {author} {\bibfnamefont {B.}~\bibnamefont {Lorenz}}, \bibinfo {author}
  {\bibfnamefont {F.}~\bibnamefont {Chen}}, \bibinfo {author} {\bibfnamefont
  {A.}~\bibnamefont {Guloy}}, \ and\ \bibinfo {author} {\bibfnamefont
  {C.}~\bibnamefont {Chu}},\ }\href@noop {} {\bibfield  {journal} {\bibinfo
  {journal} {New Journal of Physics}\ }\textbf {\bibinfo {volume} {11}},\
  \bibinfo {pages} {025013} (\bibinfo {year} {2009})}\BibitemShut {NoStop}%
\bibitem [{\citenamefont {Hodovanets}\ \emph {et~al.}(2011)\citenamefont
  {Hodovanets}, \citenamefont {Mun}, \citenamefont {Thaler}, \citenamefont
  {Budko},\ and\ \citenamefont {Canfield}}]{Hodovanets11}%
  \BibitemOpen
  \bibfield  {author} {\bibinfo {author} {\bibfnamefont {H.}~\bibnamefont
  {Hodovanets}}, \bibinfo {author} {\bibfnamefont {E.~D.}\ \bibnamefont {Mun}},
  \bibinfo {author} {\bibfnamefont {A.}~\bibnamefont {Thaler}}, \bibinfo
  {author} {\bibfnamefont {S.~L.}\ \bibnamefont {Budko}}, \ and\ \bibinfo
  {author} {\bibfnamefont {P.~C.}\ \bibnamefont {Canfield}},\ }\href@noop {}
  {\bibfield  {journal} {\bibinfo  {journal} {Phys. Rev. B}\ }\textbf {\bibinfo
  {volume} {83}},\ \bibinfo {pages} {094508} (\bibinfo {year}
  {2011})}\BibitemShut {NoStop}%
\bibitem [{\citenamefont {Liu}\ \emph {et~al.}(2011)\citenamefont {Liu},
  \citenamefont {Palczewski}, \citenamefont {Dhaka}, \citenamefont {Kondo},
  \citenamefont {Fernandes}, \citenamefont {Mun}, \citenamefont {Hodovanets},
  \citenamefont {Thaler}, \citenamefont {Schmalian}, \citenamefont {Bud'ko},
  \citenamefont {Canfield},\ and\ \citenamefont {Kaminski}}]{Liu11}%
  \BibitemOpen
  \bibfield  {author} {\bibinfo {author} {\bibfnamefont {C.}~\bibnamefont
  {Liu}}, \bibinfo {author} {\bibfnamefont {A.~D.}\ \bibnamefont {Palczewski}},
  \bibinfo {author} {\bibfnamefont {R.~S.}\ \bibnamefont {Dhaka}}, \bibinfo
  {author} {\bibfnamefont {T.}~\bibnamefont {Kondo}}, \bibinfo {author}
  {\bibfnamefont {R.~M.}\ \bibnamefont {Fernandes}}, \bibinfo {author}
  {\bibfnamefont {E.~D.}\ \bibnamefont {Mun}}, \bibinfo {author} {\bibfnamefont
  {H.}~\bibnamefont {Hodovanets}}, \bibinfo {author} {\bibfnamefont {A.~N.}\
  \bibnamefont {Thaler}}, \bibinfo {author} {\bibfnamefont {J.}~\bibnamefont
  {Schmalian}}, \bibinfo {author} {\bibfnamefont {S.~L.}\ \bibnamefont
  {Bud'ko}}, \bibinfo {author} {\bibfnamefont {P.~C.}\ \bibnamefont
  {Canfield}}, \ and\ \bibinfo {author} {\bibfnamefont {A.}~\bibnamefont
  {Kaminski}},\ }\href {\doibase 10.1103/PhysRevB.84.020509} {\bibfield
  {journal} {\bibinfo  {journal} {Phys. Rev. B}\ }\textbf {\bibinfo {volume}
  {84}},\ \bibinfo {pages} {020509} (\bibinfo {year} {2011})}\BibitemShut
  {NoStop}%
\end{thebibliography}%

\end{document}